\documentstyle[onecolumn]{mn}
\newcommand{\mincir}{\raise -2.truept\hbox{\rlap{\hbox{$\sim$}}\raise5.truept
\hbox{$<$}\ }}
\newcommand{\magcir}{\raise -2.truept\hbox{\rlap{\hbox{$\sim$}}\raise5.truept
\hbox{$>$}\ }}
\newcommand{\minmag}{\raise-2.truept\hbox{\rlap{\hbox{$<$}}\raise 
6.truept\hbox
{$>$}\ }}
\newcommand{\be}{\begin{equation}}
\newcommand{\ee}{\end{equation}}
\newcommand{\ba}{\begin{eqnarray}}
\newcommand{\ea}{\end{eqnarray}}
\newcommand{\brr}{\begin{array}}
\newcommand{\err}{\end{array}}
\newcommand{\bc}{\begin{center}}
\newcommand{\ec}{\end{center}}
\newcommand{\br}{\mbox{\bf r}}
\newcommand{\bv}{\mbox{\bf v}}

\def\ifm#1{\relax\ifmmode#1\else$\mathsurround=0pt #1$\fi}
\def\kms{\ifmmode\,{\rm km}\,{\rm s}^{-1}\else km$\,$s$^{-1}$\fi}
\def\hmpc{\ifmmode\,{\it h }^{-1}\,{\rm Mpc }\else $h^{-1}\,$Mpc\,\fi}

\def\fig #1, #2, #3 {
  \smallskip
  \centerline{\psfig{figure=#1,height=#2 in,width=#3 in}} }

\def\\{\hfill\break}

\def\etal{{\it et al.\ }}

\def\PZ{PSC$z$\ }

\def\ifm#1{\relax\ifmmode#1\else$\mathsurround=0pt #1$\fi}
\def\kms{\ifmmode\,{\rm km}\,{\rm s}^{-1}\else km$\,$s$^{-1}$\fi} 

\def\ltsima{$\; \buildrel < \over \sim \;$}
\def\lsim{\lower.5ex\hbox{\ltsima}}
\def\gtsima{$\; \buildrel > \over \sim \;$}
\def\gsim{\lower.5ex\hbox{\gtsima}}
 
\def\pmb#1{\setbox0=\hbox{#1}%
 \kern-.025em\copy0\kern-\wd0
 \kern.05em\copy0\kern-\wd0
 \kern-.025em\raise.0433em\box0}
\def\vv{\pmb{$v$}}

\def\v0{\pmb{$0$}}
\def\vnabla{\pmb{$\nabla$}}
\def\div{\vnabla\!\cdot\!}
\def\div{\vnabla\!\cdot\!}

\title [PSCz Velocity Field]{A Non Parametric Model for the Cosmic
Velocity Field}
\author[E. Branchini \etal]
{E. Branchini$^{1}$, L. Teodoro$^{1}$, C.S. Frenk$^{1}$, I. Schmoldt$^{2}$, \\
\vspace{-1mm}\\ 
{\LARGE G. Efstathiou$^{3}$, S.D.M. White$^{4}$ 
W. Saunders$^{5}$, W. Sutherland$^{2}$,   }\\
\vspace{-1mm}\\ 
{\LARGE M. Rowan-Robinson$^{6}$,   O. Keeble$^{6}$, H. Tadros$^{7}$, 
S. Maddox$^{3}$, S. Oliver$^{6}$, }\\ 
$^1$Department of Physics, University of Durham, South Road, Durham DH1
 3LE, UK \\
$^2$ Department of Physics, University of Oxford, Keble Road, Oxford
OX1 3RH, UK \\
$^3$ Institute of Astronomy, University of Cambridge, Madingley Road,
Cambridge CB3 OHA, UK\\
$^4$ Max Planck Institut f\"{u}r Astrophysik,
Karl-Schwarzschild-Stra{\ss}e 1, 85740 Garching, Germany \\
$^5$ Institute for Astronomy, University of Edinburgh, Blackford Hill,
Edinburgh EH9 3JS, UK  \\
$^6$ Imperial College of Science, Technology, and Medicine, Blackell
Laboratory, Prince Consort Road, London SW1 2EZ, UK\\
$^7$ Department of Physics, University of Sussex, Falmer, Brighton BN1
9QH, UK }

\begin{document}

\maketitle
\begin{abstract}

We present a self--consistent nonparametric model of the local cosmic velocity
field derived from the distribution of IRAS galaxies in the \PZ redshift
survey. The survey was analyzed using two independent methods, both based on
the assumptions of gravitational instability and linear biasing. The two
methods, which give very similar results, were tested and calibrated on mock
\PZ catalogues constructed from cosmological N-body simulations.  The denser
sampling provided by the \PZ survey compared to previous IRAS galaxy surveys
allows an improved reconstruction of the density and velocity fields out to
large distances. The most striking feature of the model velocity field is a
coherent large--scale streaming motion along the baseline connecting
Perseus--Pisces, the Local Supercluster, the Great Attractor, and the Shapley
Concentration. We find no evidence for
back-infall onto the Great Attractor. 
Instead, material behind and around the Great Attractor is inferred to be
streaming towards the Shapley Concentration, aided by the compressional
push of two large nearby underdensities. The \PZ model velocities compare
well with those predicted from the 1.2Jy redshift survey of IRAS galaxies
and, perhaps surprisingly, with those predicted from the distribution of
Abell/ACO clusters, out to 140 $h^{-1}$ Mpc.  Comparison of the real-space
density fields (or, alternatively, the peculiar velocity fields) inferred
from the \PZ and cluster catalogues gives a relative (linear) bias
parameter between clusters and IRAS galaxies of $b_c=4.4\pm0.6$. Finally,
we implement a likelihood analysis that uses all the available information
on peculiar velocities in our local universe to estimate
$\beta=\Omega_0^{0.6}/b=0.6^{+0.22}_{-0.15}$ (1--$\sigma$), where $b$ is
the bias parameter for IRAS galaxies.
\end{abstract}


\newpage
\section{Introduction}
\label{sec:intro}

The art of modelling the cosmic velocity field, which originates from the
desire to interpret observed deviations from a uniform Hubble expansion,
has developed rapidly over the past few years. There are two main reasons
for this. One is an increase in the quantity and quality of measured
galaxy peculiar velocities. The other is the advent of nearly all--sky,
flux--limited, redshift surveys that allow self--consistent
theoretical predictions to be made with the requisite accuracy.

Although other possibilities have been proposed (e.g. Babul \etal 1994),
the gravitational instability theory (Peebles 1980) has proven to be the
most successful theoretical framework in which to interpret peculiar
velocity data in relation to inhomogeneities in the mass distribution.
Early attempts to account for observed velocities within this general
framework were rather simplistic due to incomplete knowledge of the
distribution of galaxies in the local universe. Thus, simple parametric
models were developed to describe cosmic flows in terms of infall onto
one or more spherical overdensities such as the Virgo Cluster, the Great
Attractor or the Perseus Pisces supercluster (Davis and Peebles 1983,
Lynden--Bell \etal 1987, Peebles 1988, Han and Mould 1990). The situation
changed dramatically when statistically complete, nearly all--sky, redshift
catalogues of galaxies were constructed since it then became possible to
predict peculiar velocities directly, assuming that luminous objects trace
the underlying density field in some fashion. Since then, several
modelling procedures have been developed which are generally based on the
simplifying assumption that the gravitating mass is distributed just like
the tracer objects (galaxies or clusters), although the relative
amplitude of the deviations from uniformity,
usually called  the bias, is taken to be a free parameter. In
addition to this ``linear bias model," current methods also assume
gravitational instability in the linear or mildly nonlinear regime (e.g.
Yahil \etal 1991, Kaiser \etal 1991, Nusser and Davis 1994, Fisher et
al. 1995b, Sigad \etal 1998), and so require smoothing over scales
where non-linear effects are important. This necessitates the additional
assumption that smoothing a distribution that has evolved to a
nonlinear state gives a result that can be modeled by linear or 
quasilinear evolution from smoothed initial conditions.

Because of their large sky coverage, the most extensively used redshift
surveys are those based on the ``Point Source Catalogue" (PSC) of IRAS
galaxies (e.g. the 1.9 Jy survey of Strauss \etal 1990, the upgraded 1.2 Jy
catalogue of Fisher \etal 1995a, or the deeper but sparser QDOT survey of
Rowan--Robinson \etal 1990). Other catalogues containing different kinds of
objects such as optically selected galaxies (e.g. Shaya, Tully and Pierce
1990, Hudson 1994, Baker \etal  1998) or clusters of galaxies (Scaramella 1995
Branchini and Plionis 1996) have also been used to  produce model  velocity fields on
scales up to 200 \hmpc\footnote{ Throughout this paper we write 
Hubble's constant as $H_0=100 h\ {\rm km\ s}^{-1}\ {\rm Mpc}^{-1}$}. 

Comparison of a model velocity field with directly measured peculiar
velocities provides a means to constrain the density parameter, $\Omega_0$,
while also testing the gravitational instability hypothesis and the assumed
biasing scheme. A successful model for the peculiar velocity field may be
used to recover the distribution of galaxies or clusters in real space,
directly from their measured redshifts. This, in turn, allows investigation
of the statistical properties of the objects' distribution, free from the
effects of redshift space distortions.  Overall, model velocity fields
constructed using different surveys have proved to be remarkably consistent
with one another (Freudling, da Costa and Pellegrini 1994, Baker et
al. 1998) and have succeeded in reproducing most of the characteristics
exhibited by maps made directly from observed peculiar velocities.  There
are, however, two notable exceptions: the large bulk motion on very large
scales claimed by Lauer and Postman (1994) and a dipolar coherence in the
residuals between the velocity field obtained from the Mark III catalogue
(Willick \etal 1997) and the IRAS 1.2Jy gravity field on scales of $\sim
$50 \hmpc (Davis, Nusser and Willick 1996).

Here we present a new nonparametric model of the cosmic velocity field
based on the recently completed \PZ survey of IRAS galaxies. This is the
last of the nearly all-sky redshift surveys based on the IRAS catalogue and
supersedes both the 1.2Jy and the QDOT catalogues which it contains. The
denser sampling and lower flux limit of the \PZ survey allows us to model
the peculiar velocity field on larger scales than was possible before
without excessive contamination from shot noise.

The outline of this paper is as follows. In \S 2 we describe the \PZ
dataset as well as two other redshift catalogues that we use to construct
independent model velocity fields. Two methods for generating the \PZ
peculiar velocity fields have been implemented in order to keep track of
systematic errors. These are presented in \S 3, together with a detailed
error analysis. A cosmographic tour is performed in \S 4, along with a
consistency check between the two \PZ velocity models and a comparison
between the \PZ and 1.2Jy model velocity fields. In \S 5 we take
advantage of the large depth of the \PZ survey to compare the gravity
field derived from it with the one derived from the distribution of
Abell/ACO clusters. An estimate of the parameter $\beta \simeq
\Omega_0^{0.6}/b$, where $b$ is the linear bias parameter of IRAS galaxies,
is obtained in \S 6 by comparing observed and predicted bulk velocity
vectors. In \S 7 we discuss our results further and summarise our main
conclusions.

\section{The Datasets}
\label{sec:dataset}

The main dataset used in this work is the recently completed \PZ redshift
survey described in detail by Saunders \etal (1998a). The catalogue
contains some 15,500 IRAS PSC galaxies with 60 $\mu m $ flux, $f_{60}>0.6$.
The construction of the catalogue is described in detail in Saunders \etal
(1998a) and so we only summarize here its main features. The selection of
potential galaxies from the PSC emphasised completeness at the expense of
contamination by cirrus, stars, AGN, etc. Because the completeness of the
PSC is not guaranteed to 0.6Jy in 2HCON areas (Beichman \etal 1988, PSC
explanatory supplement), we added to the PSC any 1HCON detections from the
Point Source Catalogue Reject file with FSS counterparts (Moshir etal
1989). We then selected sources with $f_{100} < 4 f_{60}$ (to eliminate the
vast majority of cirrus sources) and $f_{60} > 0.5 f_{25}$ (to eliminate
the vast majority of stars).  Residual contamination was eliminated by a
combination of optical identification on Schmidt plates, examination of the
raw IRAS addscan profile, information from the Simbad database and, where
still unclear, optical spectroscopy. Large galaxies can have their flux
underestimated by the IRAS beam, so we constructed an all-sky list of
optically-selected galaxies with extinction-corrected diameters larger than
$2.25'$. The PSC detections for these galaxies were flagged for
deletion. The optical catalogue was addscanned by IPAC, and we used
software provided by Amos Yahil to derive fluxes from the addscan profiles,
and where the flux was sufficient, added them to the catalogue.
 
For our purposes, one of the most important properties of the \PZ
catalogue is its large sky coverage. The only excluded regions are two
thin strips in ecliptic longitude that were not observed by IRAS, 
the Magellanic clouds, and the area in the galactic plane where
the B--band extinction,  $A_B$, exceeds 2 magnitudes. The extinction is
estimated from the 100 $\mu m $ background, and includes a correction for
the estimated temperature gradient in the Milky Way (Saunders \etal 1998a). 
Overall, the \PZ catalogue covers $\sim 84 \%$ of the sky.
Although we will occasionally consider galaxies with recession velocity as
large as 30000 \kms, for the most part we will restrict our analysis to 
the \PZ subsample of 11206 galaxies  
within 20000 \kms in the Local Group (LG) frame. The distribution on the sky
of \PZ galaxies in shown in galactic coordinates (Aitoff projection) in
the upper part of Fig.~1.  

For comparison purposes, we have also applied our analysis to the similar,
but shallower, IRAS 1.2Jy redshift survey (Fisher \etal 1995a). Galaxies
in this catalogue were also selected from the IRAS PSC but using a higher
(ADDscan) flux limit, $f_{60}>1.2$ Jy, and different criteria for
minimizing contamination by galactic cirrus.
This catalogue has a somewhat larger sky
coverage of $\sim 88 \%$. Here we will use the 4939 IRAS 1.2 Jy galaxies
within 20000 \kms of the Local Group.

Finally, in an attempt to extend our analysis, we 
use a completely different set of 
mass tracers consisting of
a volume--limited sample of optically--selected galaxy clusters
extracted from the Abell (1958) and the Abell, Corwin \& Olowin (1989;
hereafter ACO) catalogues. These were combined into a single homogeneous 
catalogue 
using the clusters in common between the two samples, according to the
prescription described by Branchini and Plionis (1996; hereafter BP96). The
sample has a limiting depth of 250 \hmpc and contains 493 clusters of
richness class $R\ge 0$  at $|b| \ge 13^{\circ}$, and $m_{10}<17$, where
$m_{10}$ is the magnitude of the tenth brightest galaxy in the cluster.

The number of galaxies in a flux limited sample decreases with distance,
as may be seen in the lower plot of
Fig.~2, for the \PZ (upper histogram) and 1.2 Jy
(lower, shaded histogram) surveys. 
We define a selection function,
$\phi(r)$, as the fraction of galaxies that can be seen out to a redshift
distance $r={{cz}\over{H_{\circ}}}$ (expressed in \hmpc), given the survey
flux limit. To determine $\phi(r)$ we
use the parametric maximum likelihood estimate proposed by 
Yahil \etal (1991), in which the following analytic form for the
selection function is assumed:
\begin{equation} 
\phi(r)=A r^{-2\alpha}\left(1+{r}^2\over{r_{\star}^2}\right)^{-\beta}
\ {\rm if}  \ r>r_{s}.
\label{eq:selection2}
\end{equation}
The value of the normalization constant, $A$, does not affect 
the modeling of peculiar velocities. In this paper we arbitrarily set
$\phi(r)=1 $ if $r\le r_s$ with $r_s=6$ \hmpc. 
This choice is equivalent to imposing a lower
cutoff in the 60 $\mu m $ luminosity which, in turn, avoids giving
too much weight to faint, nearby IRAS galaxies that may 
not trace the galaxy distribution reliably in the nearby volume 
(Rowan Robinson \etal 1990).
The relevant parameters have been determined via likelihood
analysis using only the  objects within 100 \hmpc. The results are 
displayed in  Table~1 for the \PZ and
1.2Jy  surveys, both in redshift (z-) and real (R-) space (i.e. after
correction for redshift space distortions as discussed in \S 3.1).
The number density of galaxies as a function of redshift distance,
predicted by eqn.~(1), is shown in the lower part of Fig.~2, superimposed
on the observed $N(r)$ histograms. The theoretical curves provide a good
description of the data.  The ratio of the number of observed to the number
of expected galaxies gives the radial overdensity field, shown in the upper
part of Figure~2 for both the \PZ (thick line) and the IRAS 1.2Jy
catalogues.  A horizontal line is drawn at the mean density. The radial
density fields traced by the two surveys are consistent with one another,
but the larger shot noise in the 1.2 Jy sample makes the
amplitude of density fluctuations
larger than in the \PZ survey at large radii.

Our estimator for the selection function is independent of clustering but
requires prior knowledge of the evolutionary rate. Springel and White
(1998) have recently developed a new technique to estimate the rate of
evolution of the selection function. For the 1.2Jy catalogue they find
rather strong evolution and this is even more dramatic in the \PZ
catalogue (Springel 1996). Ignoring this effect could introduce spurious
streaming motions in the model velocity field. However, we have verified
that our selection function for \PZ is very similar to the one derived by
Springel (1996) within the scales relevant to our analysis. The difference
between the two selection functions increases with distance but it is only
5\% at 200 \hmpc. Strong evolutionary effects become important only beyond
these scales and we shall therefore ignore them in our modeling.

For the Abell/ACO composite sample we adopt the selection function
derived by BP96: 
\begin{equation}
\phi(r)= \left\{ \brr{lll}
        1  & \mbox{ if $r \leq r_{c1}$} \\
        0.5(1+A_1 \exp(-r/r_{\circ1})) & \mbox{ if $r_{c1}<r\leq r_{c2}$}\\
        0.5(A_2\exp(-r/r_{\circ2})+A_1\exp(-r/r_{\circ1})) & 
        \mbox{ if $r>r_{c2}$}
        \err \right.
\label{eq:selectionc}
\end{equation}
The estimated values of the parameters are listed in Table~2, and the
expected number of clusters as a function of redshift distance is shown in
the lower plot of Fig.~2 as a thick line. 

The mean separation, $l$, of the objects in a population of tracers limits
the intrinsic resolution with which we the population probes the underlying 
cosmic fields. Thus, $R_s=l$, is a natural smoothing length which keeps
the shot noise at a constant level.  In a
magnitude limited survey, $l$ increases with distance $r$ according to: 
\begin{equation} 
l(r) = \left[ \phi(r) \langle n \rangle \right]^{-1/3},
\label{eq:smooth}
\end{equation}
where the average number density of objects $\langle n \rangle$
has been estimated as
\begin{equation} 
\langle n \rangle= 
\frac{1}{V}\sum_{i=1}^{N_g} \phi(r_i)^{-1},
\label{eq:smooth2}
\end{equation}
and the sum extends over all $N_g$ objects contained within the volume 
sampled, $V$. The value of the average number density  within
100 \hmpc, computed in redshift and real space,
is shown in Table~1 both for the \PZ and IRAS 1.2 Jy catalogues.
The more physically meaningful values of the mean galaxy separation at 100 \hmpc, $l(r)$ expressed in
$h^3$ Mpc$^{-3}$, are also given in the table. Because of our 
normalization ($\phi(r)=1$ within 6 \hmpc), 
our estimate of $\langle n \rangle$ for the 1.2 Jy galaxies is slightly 
different from that of Fisher \etal (1995a).
Fig.~3 shows the mean interobject separation  as a function of $r$ for the
three samples considered. Because of the shallower
depth of the 1.2Jy sample, its intergalaxy separation increases much more
steeply with distance than in the \PZ sample. Thus, at a fixed resolution,
the \PZ survey probes cosmic structures out to larger distances than 
the 1.2 Jy survey without an increase in shot noise. The
dot--dashed line in Fig.~3 shows the Abell/ACO mean intercluster
separation. Locally, this is much larger than the corresponding separation
in the galaxy surveys, but on scales above 170 \hmpc, clusters
start to become more effective than IRAS 1.2Jy galaxies in tracing the
cosmic density fields.

\section{Non--parametric Models of the Cosmic Velocity Field} 
\label{sec:reconstruction}

The main aim of this paper is to obtain a self--consistent, non--parametric
model of the velocity field in the local universe (i.e. for $r <150$
\hmpc). We do this by removing the redshift space distortions that affect
the observed radial galaxy distribution in the  survey using two different
procedures. The first is the iterative technique devised by Strauss \&
Davis (1988), applied by Yahil \etal (1991), and further refined by Sigad
\etal (1998). The results presented here were obtained by applying  the
original Yahil \etal (1991) technique. The second procedure, used here
primarily as a check for possible systematic effects, is the
non--iterative technique developed by Nusser and Davis (1994).

\subsection{The methods} 

The two procedures that we have implemented rely on three important
assumptions. Firstly, cosmic structures are assumed to have grown by 
gravitational amplification of small amplitude fluctuations present in the
density field at early times (gravitational instability, c.f. Peebles
1980). Secondly, fluctuations on the scales of interest are assumed to be 
small enough that linear (or mildly non-linear) theory is
applicable. Thirdly, luminous objects (galaxies or clusters) are assumed 
to trace the underlying mass density field according to: 
\begin{equation}
 \delta_l(\br)=b_l\delta(\br), 
\end{equation}
where $\delta$ is the mass overdensity at position $\br$, $\delta_l$ is
the fluctuation in the number of luminous objects at this same location,
and $b_l$ is the biasing parameter. Eqn.~(5) is often referred to as the
``linear biasing model." The reconstruction methods return a model of the
cosmic velocity field that depends on the parameter
$\beta=\Omega_0^{0.6}/b_l$. The value of $\beta$ must be determined {\it a
posteriori} by comparison with other observational data.

The two reconstruction techniques that we use are extensively discussed
and tested by Branchini \etal (1998) where the reader may find further
details. Here, we simply outline the main principles of each method.

\noindent {\bf Method 1: The iterative technique} 

In linear theory the overdensity is related to the peculiar velocity field by
\begin{equation}
\div{v} =  - \beta \delta_l.
\label{eq:vdr}
\end{equation}
The model velocity field is then obtained by iteratively solving the
system of equations:
\begin{equation}
{\bf v}({\bf r}) = \frac{H_0 \beta}{4 \pi} \int d^{3} {\bf r \prime}
\frac{{\bf r\prime} - {\bf r}}{\mid {\bf r\prime} - {\bf r} \mid^{3}}
\delta_l({\bf r\prime}) 
\label{eq:vir}
\end{equation}
and
\begin{equation}
{ r} = cz - \hat{\bf r} \cdot ({\bf v} - {\bf v}_{LG}). 
\label{eq:czr}
\end{equation} 
Eqn (\ref{eq:vir}) follows from integrating eqn~(\ref{eq:vdr}); ${\bf v}$ is the
peculiar velocity of a tracer object at position given by the radial unit
vector $\hat{{\bf r}}$; and ${\bf v}_{LG}$ is the velocity of the Local
Group.

During each iteration, the selection function and the mean number density
of the population of displaced objects are recomputed. Eqns~(\ref{eq:vdr})
and~(\ref{eq:vir}) are only valid in the linear regime and so the force
field needs to be smoothed to eliminate nonlinear effects. We employ
a ``top hat" filter with a smoothing radius $R_s(r)=max[5,(\langle n
\rangle \phi(r))^{-1/3}]$ \hmpc. This choice eliminates most of the
nonlinear contributions and keeps the shot noise at a constant level. To
improve numerical convergence, we adiabatically increase the value of
$\beta$ at each of ten iterations, from 0.1 to 1.0 (Strauss \etal 1992b).
Note that the amplitude of the velocity vectors scales, to first order,
linearly with $\beta$. Unless otherwise stated, all the plots of the model
velocity fields that we present assume $\beta=1$.

Around high density regions, such as clusters, there may be triple--valued
regions in which the same redshift is observed at three different
positions along a given line-of-sight. We correct for this by collapsing
the galaxies within clusters and implementing the ``robust procedure"
(``method~2") of Yahil \etal (1991) to determine the locations of galaxies
within triple--valued regions. For the collapsing procedure, we have used
two different datasets:  the six nearest clusters listed in Table~2 of
Yahil \etal  (1991) and a larger catalogue of 61 objects obtained by merging
the 59 SMAC clusters of Hudson \etal (1998) with the Yahil \etal  (1991)
dataset. The results show that the collapsing procedure only affects the
infall pattern around rich clusters. However, with the level of smoothing
used here, the effect is negligible and no systematic differences are
found when the model velocity fields are compared on a point--by--point
basis.

\noindent {\bf Method 2: Spherical harmonics expansion} 

This procedure is based on the method proposed by Nusser and Davis
(1994). In linear theory, the velocity field in redshift space is
irrotational and thus may be derived from a velocity potential:
\begin{equation}
{\bf v}({\bf s})=-\nabla \Phi({\bf s}).
\label{eq:irr}
\end{equation}
Expanding the potential and the galaxy density field,
$\delta^g$, in spherical harmonics, and using the Zel'dovich approximation,
it can be seen that the two fields obey a Poisson--like equation:
\begin{equation}
{{1}\over{s^2}}{{d}\over{ds}}
\left( s^2 {{d\Phi_{lm}}\over{ds}} \right)
-{{1}\over{1+\beta}}
{{l(l+1)\Phi_{lm}}\over{s^2}}=
{{\beta}\over{1+\beta}}
\left( \delta^g_{lm}-
{{1}\over{s}}{{d\ln{\phi}}\over{d \ln{s}}}
{{d \Phi_{lm}}\over{ds}} \right),
\label{eq:nd}
\end{equation}
where $\delta^g_{lm}$ and $\Phi_{lm}$ are the spherical harmonic
coefficients of the galaxy density and velocity potential fields
respectively. We have considered the coefficients up to 
$l \le l_{max}=15$ and $|m| \le l_{max}$.
We employ a Gaussian filter with smoothing radius 
$R_s(r)=max[3.2,(\langle n \rangle \phi(r))^{-1/3}]$ \hmpc,
compute the coefficients $\delta^g_{lm}$ in redshift
shells for the different catalogues, solve eqn.~(\ref{eq:nd}) for $\Phi_{lm}$,
and then compute the full three--dimensional velocity field in redshift
space using eqn.~(\ref{eq:irr}). Galaxy positions and peculiar velocities at
the real space positions, ${\bf r}$, are obtained by assuming a one--to--one
mapping between the real and redshift space positions, ${\bf r}$ and 
${\bf s}$, along the line-of-sight. We
minimize the problem of triple--valued regions by adopting the same
cluster--collapsing scheme used above. 

Redshift space distortions are only one of several effects that hamper the
recovery of cosmic density and velocity fields from observational data.
Incomplete sky coverage is another, potentially serious, problem and we
deal with it using a filling method similar to that introduced by Yahil
\etal (1991). We use their original technique to fill in the region at
galactic latitude $|b|\le8^{\circ}$, but each real \PZ galaxy at $|b|\le
8^{\circ}$ is included to replace a synthetic object that resides in the
same  longitude--distance bin. Masked regions at larger galactic latitudes
are filled in with a random distribution of synthetic galaxies having the
observed mean number density. Tests performed by Branchini \etal (1998)
using the mock catalogues described in the next section show that 
the only bias introduced by this filling in technique is a 
spurious bulk motion of $< 60$ \kms (see \S 6.1).
As a further check we have also used the new
Fourier interpolation technique developed by Saunders et al. (1998b),
which interpolates structure across the mask under the assumption that the
density field of galaxies is a lognormal random field. In doing this, we
have used a `tapered', or radially-dependent, mask to take into account the
known incompleteness in the \PZ at large distances ($> 15000$ \kms) and
low latitudes ($A_B>1^m$), due to difficulties in making secure
identifications and obtaining redshifts for obscured distant galaxies. The
area of sky across which we interpolate is defined by $A_B = 2^m$ at
$V<15000 \kms$, increasing to $A_B = 1^m$ at $V=30000 \kms$. Outside this
volume, the \PZ is essentially complete. The model velocities predicted
using this mask turned out to be almost identical to the ones obtained
using the simpler filling in procedure described above.

Another source of random and systematic error is the radial selection
function which, when coupled to redshift distortions, generates the so
called 'rocket effect' discussed by Kaiser and Lahav (1989). This is a 
spurious acceleration measured from a magnitude-limited sample of galaxies
by an observer who has a peculiar velocity unrelated to the true 
gravitational acceleration. 
In Method~1 this effect is quantified and corrected for
using the mock catalogues discussed below.
In Method~2 the rocket effect is explicitly 
accounted for by the second term in the right-hand side of eqn.~(\ref{eq:nd}).

Finally, another potential source of bias is the fact that IRAS galaxies
are preferentially of late morphological type and are less clustered than
early types (Strauss \etal 1992a, Hermit \etal 1996). However, any biases
arising from this are likely to be small since Baker \etal (1998) have
shown that the ratio of the amplitudes of IRAS and optical galaxy density
fields, smoothed on scales $\ge 5 \hmpc$, is almost constant, implying a
ratio of linear biasing parameters of $b_o/b_I\simeq 1.4$.

\subsection{Error estimates using mock catalogues} 
\label{sec:mock1}

We have used a suite of large cosmological N-body simulations to generate
mock galaxy catalogues that mimic the main properties of the \PZ and 1.2Jy
redshift surveys. We use these to quantify random and systematic errors in
our reconstructed velocity fields. For the fields inferred from cluster
catalogues, we adopt the error estimates derived by BP96 and by Branchini
\etal (1997) on the basis of a hybrid Monte Carlo/mock catalogue technique.

The simulations we use are 
those described by Cole \etal (1998). We consider two different cold dark
matter cosmologies: a flat model with $\Omega_0=0.3$ and cosmological
constant term, $\Lambda c^2/3H_0^2=0.7$, and a critical density universe
($\Omega_0=1.0$) with power spectrum shape parameter, $\Gamma=0.25$. In both
cases, the fluctuation amplitude was normalized by the observed abundance
of galaxy clusters. This requires setting $\sigma_8=0.55 \Omega_0^{-0.6}$
(Eke, Cole \& Frenk 1996), where $\sigma_8$ is the linear rms mass density
fluctuation in top hat spheres of radius 8\hmpc. Galaxies were identified
with random particles from the simulations (so that $b=1$ by
construction). Prior to generating the mock catalogues, we smoothed
the velocities in the simulations using a top hat  filter of radius 1.5
\hmpc. This brings $\sigma_{12}(1)$, the pairwise velocity dispersion
of objects with projected separation $\le 1$ \hmpc, down to 
$\sim 250 $ \kms (when only particles within 30 \hmpc from the
observer are considered). This value is in accordance with the recent analysis
of the Optical Redshift Survey of Santiago \etal (1995, 1996) by 
Strauss, Ostriker and Cen (1998) for galaxies outside high density regions.
and with  the estimate of $\sigma_{12}(1)$ for late type galaxies 
in the Perseus--Pisces redshift catalogue by Guzzo \etal (1997).
Both these measurements refer to `field' galaxies, and are therefore
well matched to the late type IRAS galaxies of the \PZ catalogue.

Each mock \PZ catalogue is contained in a sphere of radius 170 \hmpc and,
for the purposes of error analyses, 10 nearly independent mock catalogues 
were obtained from each cosmological model. Several constraints were
applied in order to obtain mass distributions and velocity fields as
similar as possible to those observed:

\noindent 
1) Local Group observers. Hypothetical observers were selected from the set
of particles with velocity, ${\bf v}_{LG}= 625 \pm 25$ \kms, lying 
in regions in which the shear within 5 \hmpc is smaller than 200
\kms and the fractional overdensity  within the same  scale
ranges between -0.2 and 1.0
(Brown and Peebles 1987). These constraints mimic the Local Group 
environment. 
 
\noindent
2) Coherent galaxy dipole. A galactic coordinate system, $(l,b)$, in the
(periodic) computational volume was defined such that the velocity of the
observer pointed towards $(l,b)=(276,30)$, the direction of the dipole
anisotropy observed in the cosmic microwave background (CMB) radiation.
This dipole is known to be approximately aligned with the dipole in the
distribution of IRAS galaxies (Strauss \etal 1992a, Schmoldt \etal 1998).

\noindent
3) Radial selection. We generated flux--limited ``galaxy'' samples using a
Monte Carlo rejection procedure to select particles in the simulations,
assigning them fluxes according to eqn.~(1). In the vicinity of an observer,
the particle number density in the simulation is less than the number
density of galaxies in the \PZ and 1.2Jy catalogues. We therefore
generated catalogues that are semi--volume limited at a radius of
10.9 \hmpc for \PZ and 7.8 \hmpc for 1.2Jy.

\noindent
4) Masked areas. To mimic the incomplete sky coverage, we excluded 
all objects within the unobserved regions in the two IRAS catalogues.

The distribution on the sky of the galaxies in one of our \PZ mock
catalogues is shown, as an Aitoff projection, in the lower panel of
Fig.~1, where it may be compared to the real survey displayed in the upper
panel. An illustration of the utility of these mock catalogues is provided in
Fig.~4. The upper left--hand panel displays a two dimensional slice,
corresponding to the mock Supergalactic plane, through the density and
velocity fields within a region of 120 \hmpc. Both fields were tabulated on
a $64^3$ grid and subsequently smoothed using a Gaussian filter with a 
constant effective radius of 6 \hmpc. 
The upper right--hand panel shows density and velocity fields
reconstructed using Method~1. All the main features, as well as most of
the small scales structure, are correctly reproduced. The agreement
between the fields can be better appreciated from the two lower plots. The
one on the left shows the absolute value of the errors in the
reconstructed density field.  The main discrepancies occur within the 
zone-of-avoidance, depicted
as a shaded region, and at
large distances where the shot noise is large. A correlation between the
error and the signal is present only in the outer regions. The lower
right--hand plot displays the differences between the two velocity vector
fields. Again, the largest discrepancies occur at large distances and no
strong coherence is detected. Very similar results are obtained when the
reconstruction is performed using Method~2.

Unless otherwise stated, the same Gaussian filter of 6 \hmpc is used throughout the
rest of this paper. As discussed in \S 2, a smoothing radius of 6 \hmpc
corresponds to the average galaxy separation at $\sim80 \hmpc$, i.e to the
maximum distance out to which comparisons with observed peculiar
velocities are still possible. The drawback of using a constant smoothing
is that the shot noise increases with distance, artificially enhancing the
density contrast in the outer regions, as may be seen in Fig.~4. With our
adopted smoothing, the average shot noise in $\delta({\bf r})$ at a
distance of 80\hmpc is $\sim 96 \%$, rapidly increasing to $\sim 180$ \%
at 120 \hmpc. A complete description of the mock catalogues, their use for
error estimation and a detailed assessment of the reliability of our
galaxy reconstruction methods is given in Branchini \etal (1998). The
corresponding analysis for the Abell/ACO cluster catalogues may be found
in Branchini \etal (1997).

\section{A Cosmographic Tour}
\label{sec:vfc}
 
In this Section we present a qualitative description of the model density
and velocity fields of the local universe derived from the \PZ survey. We
analyse the data using Method~1 above. Visualizing three dimensional
structures and the corresponding vector fields, is not easy. Fortunately
from this point of view, the most interesting structures in the nearby
universe, on scales larger than the Local Supercluster, are roughly
distributed along a planar structure, de Vaucouleurs' (1953) Supergalactic
plane (at Supergalactic coordinates SGZ$=0$). In this analysis we will
mainly follow the custom of displaying the density and velocity features
in a two dimensional slice through this plane. However, distributions
along other parallel planes will also be displayed.
 
With the depth and sampling frequency of the \PZ survey, a reliable map
of the density field can be constructed out to a distance of 
150 \hmpc. Fig.~5 shows a slice through the Supergalactic plane of an
{\it adaptively smoothed} overdensity field. Within 30 \hmpc a constant Gaussian
filter of radius 3 \hmpc was used, but beyond that the radius of the
filter increases linearly with distance up to 11.25\hmpc at 150 \hmpc.
This  variable smoothing filter was chosen to maintain a roughly constant
sampling error over the large range of scales considered. It represents a
compromise between resolution in the inner regions and acceptable shot
noise errors near the edge. The drawback is that, unlike the case of
constant smoothing, the density contrast decreases with distance, as
is clear from the map. (The $\delta=0$ level is indicated by the yellow
line.) The most striking feature of this map is the large--scale coherence
of the mass distribution. Interconnected overdensities, separated by very
large voids, extend over distances of order 100 \hmpc.  The most
prominent structure, which plays a major role in the dynamics of the local
flow field, is the overdense ridge extending  from the Perseus--Pisces
supercluster, close to the centre of the map, all the way to the Shapley
concentration near the top left corner.
 
The \PZ survey is large enough that the velocity field corresponding to
the mass distribution can be reconstructed quite accurately, with a
relative error always smaller than 50\% in the region depicted in Fig.~5.
This precision, however, is higher than that that of measured galaxy peculiar
in a volume this size and so we cannot compare
our model predictions with velocity data over the entire region of Fig.~5.
A reliable comparison is only possible over a smaller volume, typically of
radius $\sim$80\hmpc. For this reason we now describe in some detail our
reconstructions within this distance.

\subsection{The \PZ model density and velocity fields within 
80 \hmpc} 

Fig.~6 shows the \PZ model density and velocity fields smoothed with a
6\hmpc Gaussian in a slice through the Supergalactic plane.
With this smoothing, the Local Supercluster, centered on the Virgo
cluster at ${\rm (SGX,SGY)}=(-2.5,11.5)$, does not appear as an isolated
structure but is connected instead to the prominent Hydra--Centaurus
supercluster at ${\rm (SGX,SGY)}=(-35,20)$. Together with the
Pavo--Indus--Telescopium supercluster [${\rm (SGX,SGY)}=(-40,-15)$], the latter
makes up the well known Great Attractor. The Coma cluster and its
neighbour, A1367, appear as a peak at ${\rm (SGX,SGY)}=(0,75)$, slightly
elongated in the SGX direction. The Perseus--Pisces supercluster, 
clearly visible at ${\rm (SGX,SGY)}=(45,-20)$, is the second largest peak on the
map, well separated from its northern extension [${\rm (SGX,SGY)}=(45,20)$]
which is sometimes called the Camelopardalis supercluster. Finally, the
Cetus Wall may be seen as an elongated structure around ${\rm (SGX,SGY)}=(15,-50)$.
The Sculptor void [${\rm (SGX,SGY)}=(-20,-45)$] is the largest underdensity in
the map, but is almost matched in size by a void in the background of the
Camelopardalis supercluster. Three more underdense regions that exert an
important influence on the local dynamics are the voids in the foreground
of Coma [${\rm (SGX,SGY)}=(-10,50)$], in the background of the Perseus--Pisces
complex [${\rm (SGX,SGY)}=(50,-50)$], and behind the Great Attractor
[${\rm (SGX,SGY)}=(-60,10)$]. 

The competing dynamical roles of the various overdense and underdense
structures seen in Fig~6. determine the local velocity field in a complex
manner that cannot be simply described as a bulk flow or by a multi--spherical
infall model. The local velocity field implied by the \PZ density field
is illustrated by the vectors plotted in Fig.~6. Its dominant features are
the large infall patterns towards the Great Attractor, Perseus--Pisces, Cetus Wall and
Coma. A striking property is the large-scale coherence of the velocity
field, apparent as a long ridge between Cetus and Perseus--Pisces and as a
large--scale flow along the Camelopardalis, Virgo, Great Attractor
baseline and beyond (see \S 5). A prediction of the \PZ data 
is the lack of prominent back-infall onto the Great Attractor
region. The flow around it appears to be determined by the
compressional push of two voids (at ${\rm (SGX,SGY)}=(-10,50)$ and
$(-60,10)$) and the pull of the Shapely Concentration on much larger
scales (see \S 5). All these features are also present when Method~2 is used
for the reconstruction (\S4.2), or when the 1.2Jy survey is used as the
input catalogue (\S4.3). As pointed out by Davis, Nusser \& Willick
(1996), this model velocity field does not match the Mark III peculiar
velocity field (Willick \etal 1997a) which exhibits a large outflow away
from the Centaurus supercluster and an inflow onto the Hydra complex.
There is also no evidence for a motion of the 
Perseus Pisces supercluster towards us.

Fig.~7 extends our qualitative analysis to two planes of constant
SGZ, 40\hmpc above and below the Supergalactic plane. 
The slice at SGZ$=+40$ \hmpc shows a region which is largely 
underdense (Fig~7b), dominated by a void which is connected
to the Local Void identified by Tully (1987). The prominent peak at
${\rm (SGX,SGY)}=(40,-30)$ seems to be an extension of the Perseus--Pisces
supercluster and the one at ${\rm (SGX,SGY)}=(-50,20)$ appears to be 
connected to the Pavo--Indus--Telescopium complex. 
Overall the density field traced by the \PZ survey looks remarkably
similar to the one obtained from the IRAS 1.2Jy catalogue (see \S 4.3
and the analogous discussion by Strauss and Willick 1994).
The velocity field (Fig~7a) is still characterized by a
coherent, large-scale, flow towards the same direction
[${\rm (SGX,SGY)}=(-50,50)$] as the stream seen in the Supergalactic plane. At
negative SGZ$=-40$ \hmpc the extensions of the Pavo--Indus--Telescopium
$(-50,15)$ and Perseus--Pisces $(20,-20)$ superclusters are visible. The
dynamical effect of these two peaks is evident in the associated  
large infall patterns (Fig.~7c).

\subsection{Comparison of the two reconstruction methods} 

A detailed quantitative error analysis of mock \PZ catalogues performed by
Branchini \etal (1998) also shows that the predicted velocity field is not
affected by any significant systematic bias. To check this result further
we have compared the reconstructions produced by the two different methods
discussed in \S3.1. The two upper plots in Fig.~8 show the predicted
velocity fields smoothed with a 6\hmpc Gaussian in a slice through the
Supergalactic plane. The two methods succeed in reproducing the main
features that we have already highlighted: remarkable large-scale
coherence in the velocity field, clear infall patterns onto Coma,
Perseus--Pisces and the Great Attractor but no back-infall onto the
latter. The only noticeable difference is that Method~2 seems to blur
slightly the sharp features (like the Cetus ridge) produced by Method~1 in
the velocity field.

Quantitative evidence for the similarity of the two model fields is
provided in the lower part of Fig~8, which shows a scatterplot of the SGY
Cartesian components of the reconstructed velocity fields, smoothed   with
a 6\hmpc Gaussian, measured within 80\hmpc. For both model velocity fields
 we have used $\beta=1$. In the plots we show only 1000 points taken at
random from the 73490 available. The peculiar velocities in the two models
should obey the linear equation,
\begin{equation}
M_2=B M_1+A,
\end{equation}
where $M_1$ and $M_2$ denote any of the Cartesian components 
of the peculiar velocity predicted by Methods~1 and~2, respectively, 
and $B$ is expected to be 1 if no systematic errors are present.
The offset $A$ allows for possible differences in the predicted bulk 
flows in the two models. We obtain the values of $A$ and $B$ by 
minimizing 
\begin{equation}
\chi^2=\sum_{i=1}^{N_t}{(M_{2,i}-A-B M_{1,i})^2\over(\sigma_{M_1,i}^2+B^2
\sigma_{M_2,i}^2)},
\label{eq:chi2a}
\end{equation}
where $\sigma_{M_1,i}$ and $\sigma_{M_2,i}$ are the errors in the
velocities at a generic gridpoint $i$ in the two methods, 
and $N_t$ is the total number of points used for the comparison.
We assume that 
$\sigma_{M_1,i} = \sigma_{M_2,i} \equiv \sigma $, so that 
eqn.(~\ref{eq:chi2a}) becomes 
\begin{equation}
\chi^2= {1\over \sigma^2}
\sum_{i=1}^{N_t}{(M_{2,i}-A-B M_{1,i})^2\over(1+B^2)}
={\chi^2_o\over \sigma^2}.
\label{eq:chi2b}
\end{equation}
Only $N_i$ of the $N_t$ points used in the comparison are independent. It
can be shown that the quantity $\chi^2_{eff}=(N_i/N_t)\chi^2$ is
approximately distributed as $\chi^2$ with $N_{d.o.f}=N_i-2$ degrees of
freedom (e.g. Hudson \etal 1995).  We can therefore approximately evaluate
$\sigma$ by setting $\chi^2_{eff}=N_{d.o.f}$ so that
\begin{equation}
\sigma^{2}=
{{\chi^2_o N_i}\over {N_t(N_i-2)}}.
\end{equation}
We take $N_i$ to be the number of independent volumes within the volume
sampled. With our smoothing radius, $N_i \simeq 6300$. We then find a
regression slope, B$=0.98 \pm 0.06$. The zero point, A$=-47\pm5$,
indicates a systematic difference in the SGY component of the bulk
velocity in the two models. This offset is only evident when the
two velocity fields are compared in the CMB frame and almost disappears in
the Local Group (LG) frame. As explained in \S3, the peculiar velocities
are reconstructed in the LG frame. Transformation to this frame is
performed by adding the reconstructed LG velocity at each galaxy position.
This can lead to systematic differences if the LG velocities predicted by
the two methods do not agree. Indeed, the zero point offset measured in
the velocity--velocity scatterplot for the three Cartesian components,
(-123,-47,132)$\beta$ \kms, is similar to the difference between the
predicted LG velocities in the two models, (-150,-50,98)$\beta$ \kms.
No other systematic errors are detected as the regression slopes are close
to one also for the two remaining components ($0.97\pm 0.06$ for SGX and
$1.06\pm 0.05$ for SGZ.)
For the  SGX and SGZ components, the dispersions around the fit are only
slightly larger than for SGY ($\sigma=61, \ 54,\ 70) \beta$ \kms 
respectively.) This represents the intrinsic error in the reconstruction
and is nearly a factor of 2 smaller than the average total error derived 
from the error analysis on the mock \PZ catalogues (which also accounts
for uncertainties in the filling in procedure, nonlinearity, finite volume
effects, etc) by Branchini \etal (1998).

\subsection{Comparison of the \PZ and 1.2Jy model velocity fields} 

Since they were drawn from the same parent catalogue, we expect the
\PZ and 1.2Jy catalogues to give consistent model velocity fields, at
least in the nearby volume where the sampling by 1.2Jy galaxies is not too
sparse. In analogy with the \PZ fields displayed in Fig.~6, Fig.~9 shows
the 1.2Jy density and velocity fields, smoothed with a 6\hmpc Gaussian,
in a slice through the Supergalactic plane. Most of the characteristic
structures previously identified in Fig.~6 are also visible in Fig.~9 except that 
underdense regions in the 1.2Jy map appear somewhat more extended than in
the \PZ map. The overall pattern in the \PZ velocity field is
reproduced in the 1.2Jy map although the large-scale coherent flow along
the Camelopardalis, Great Attractor baseline is less evident in the latter. 
On the other hand, the infall pattern around the Great Attractor in
the 1.2Jy survey spreads out over a larger region. These discrepancies
might well be due to the larger shot noise in the 1.2Jy catalogue.

Our 1.2Jy model velocity field is consistent with that derived
independently by Webster, Lahav and Fisher (1997) using the method
developed by Fisher \etal (1995b). Comparison with their Fig.~7d (which
assumes a CDM model with $\Gamma=0.2$ as a prior in the Wiener filtering
procedure) reveals only one noticeable difference between the two maps.
This is in the region of the Great Attractor, where the model of Webster
\etal predicts a weak back-infall. The discrepancy, however, is small and
may simply reflect our use of a constant 6\hmpc Gaussian smoothing, which
is somewhat larger than the smoothing applied by Webster \etal 

The lower part of Fig~10 displays the difference between the \PZ and the
IRAS 1.2Jy overdensity fields, displayed in Figs.~6 and 9 respectively, 
$\Delta=\delta_{PSCz}-\delta_{1.2 Jy.}$. Positive (continuous lines) and
negative (dashed lines)  contours are shown, with the thick line delineating 
the $\Delta=0$ contour. The main discrepancies between the two fields occur within the zone-of-avoidance,   
in the Great Attractor region and beyond. Outside the Galactic plane, the
radially increasing shot noise is responsible for the large discrepancies
seen in the outer regions. Apart from these local features, we notice that  
the mean value of $\Delta$, $\langle \Delta \rangle $, is positive in the
region shown in the figure. Indeed, this is true within the entire
spherical volume of radius 70 \hmpc  in which we find 
$\langle \delta_{PSCz}\rangle=7 \cdot 10^{-3}$ and  $\langle 
\delta_{1.2Jy}\rangle=-6\cdot 10^{-3}$ $h^3$Mpc$^{-3}$ 
(for both catalogues the average density has been measured within 100
\hmpc). If the errors were due to sparse sampling alone, then this
discrepancy would be significant at about the $2 \sigma$ level. However,
the errors in the estimates of the average density are larger than this. In
particular, as pointed out by Davis, Strauss and Yahil (1991), the
intimate coupling between the estimation of the selection function and the
average density itself generates uncertainties in the reconstructed
average density which are large enough to account for the discrepancy we 
have detected. The mismatch in the average densities on a scale of 70 \hmpc
generates the infall pattern visible in the upper panel of Fig~10, while
smaller scale features originate from local mismatches between the two
density fields.

Fig.~11 (lower panel) shows a more quantitative comparison between the \PZ
and 1.2Jy overdensity fields, both smoothed with a 6\hmpc Gaussian and
tabulated onto 64$^3$ grids. The comparison is made in the CMB frame and
assumes $\beta=1$. Only those gridpoints within 80 \hmpc of the LG
position are considered. As in the previous section, we fit a straight line
to the data and estimate the parameters by minimizing $\chi^2$. The slope
of this line is $1.08 \pm 0.8$, indicating that there are no systematic
differences between the two density fields. The negative zero point
offset, $-0.5 \pm 0.03$, reflects the  discrepancy  between the average
densities of the two catalogues discussed in the preceding paragraph
($\Delta \delta=-0.05$). The scatter $\sigma_{\delta}=0.20$ is similar to
the average shot noise at the gridpoint positions both for \PZ ($
\sigma_{SN}=0.18 $) and for the IRAS 1.2 Jy ($ \sigma_{SN}=0.24 $)
catalogues. Similar considerations apply to the velocity--velocity
comparison in the upper panel in which, as in Fig~8, we only show the SGY
Cartesian component of the velocity. The comparison, performed in the CMB
frame, assumes $\beta=1$. There is a non--negligible zero point offset of
$(56\pm 8) \beta$ \kms which is comparable to the difference between the
SGY components of the predicted LG velocities for two catalogues ($80
\beta$ \kms). The dispersion around the fit, $\sigma_v=104 \beta$ \kms, is
again similar to the average shot noise in the peculiar velocities,
computed as in Yahil \etal (1991) ($88 \beta$ \kms for \PZ and $140
\beta$ \kms for IRAS 1.2Jy). Similar results are found from the
scatterplots of the two remaining Cartesian components. Their slopes are
consistent with unity ($1.06 \pm 0.6$ and $0.91\pm 0.08 $ for SGX and SGZ
respectively) and the zero point offsets are comparable to the
discrepancies arising from the transformation from LG to CMB frames ($153
\beta$ and $118 \beta$ \kms).

Finally, Tadros \etal (1998) have recently suggested that the \PZ
catalogue should not be used down to 0.6 Jy and that a more conservative
cut in flux, 0.745 Jy, should be adopted instead. To check whether
possible incompleteness at the low flux limit could affect the modeling of
the \PZ velocity field we have generated several different models for
density and velocity fields using a suite of \PZ subcatalogues taken at
different flux limits. With a 0.75 Jy cut the density and velocity fields
are still remarkably similar to those with the cut at 0.6Jy. Increasing the
flux limit makes the \PZ density and velocity fields more and more like
the IRAS 1.2 Jy fields with which they almost coincide when the cut is
taken at 1.2 Jy. Overall, our prediction for the \PZ velocity and density
fields are robust and are not affected by incompleteness in the 
catalogue at low flux levels, at least within a region of 200 \hmpc.

\section{Modelling very large scale motions} 
\label{sec:vlss}

The sampling density of the \PZ survey is high enough to allow
investigation of the density field out to $\sim 120$ \hmpc even with a
6\hmpc Gaussian smoothing. This is illustrated in Fig.~12 which displays
the usual section of the density and velocity model along the
Supergalactic plane. The shot noise in this map is the same as in the maps
discussed in \S4.1. The larger volume reveals the full extent of the
coherent streaming involving galaxies from the Camelopardalis Supercluster
all the way to the Shapley Concentration (that begins to become visible at
${\rm (SGX,SGY)}=(-100,+60)$), passing through the Local Supercluster and
the Great Attractor.

The dynamical sources of the coherent motion seen in Fig.~12 are the same
ones that we identified earlier as being responsible for the flow pattern
behind the Great Attractor. The gravitational pull is mostly due to the
Shapley Concentration, but the coherence is aided by the combined push of
the two large voids (at ${\rm (SGX,SGY)}=(-80,10)$ and ${\rm (SGX,SGY)}
=(-50,70)$) that straddle the ridge and were only partially visible in
Fig.~6. This map also reveals a large extension to other underdense
regions, the Sculptor void and the two connected voids behind the
Cetus--Perseus--Pisces--Camelopardalis complex.

In Fig.~13 we show slices along four planes parallel to the Supergalactic
plane, at SGZ$=\pm 40$ and SGZ$=\pm 80$. Fig.~14 shows the corresponding
velocity fields. The same structures that we had identified in Fig.~7 can
now be traced over larger scales. The shot noise is still small and so we
can fully appreciate the large scale coherence of features in both the
density and velocity fields. For example, Figs.~13a and~13b show that the
void that we had identified in the upper half of Fig.~7b actually extends
over a much larger volume. Similarly, the stream towards
(SGX,SGY)=(-50,50) that we had identified in the SGZ=0 and SGZ=+40 \hmpc
planes persists at SGZ=+80 \hmpc. No very large scale features are present
in the plane at SGZ=-40 \hmpc (Fig.13c and 14c) which, as we had
previously noted, is dominated by a few isolated infall patterns onto the
southern extensions of the Pavo--Indus--Telescopium and Perseus--Pisces
superclusters. New large coherent structures, however, begin to appear
again in the plane at SGZ=-80 \hmpc.

\subsection{Comparison with the Abell/ACO model fields}

Various surveys of IRAS and optical galaxies have been used to construct
model velocity fields which proved to be remarkably consistent with one
another (Yahil \etal 1991, Kaiser \etal 1991, Hudson 1994, Baker \etal  
1998). We carry out a similar exercise here, using a completely different
set of mass tracers, the Abell/ACO clusters. With these we can model the density and
velocity fields out to very large scales, albeit with much larger sampling
errors. BP96 have already employed the same Abell/ACO cluster subsample
described in \S2 to model the density and velocity fields out to 250 \hmpc.
The reconstruction technique they used is a simplified version of our
Method~1 (cf. \S3.1) in which the selection function (which is nearly
constant on the scales of interest) is not iteratively updated, no special
treatment is given to triple--valued regions and the procedure for filling
in the zone-of-avoidance ($|b|\le 20^{\circ}$) is the one used by  Yahil
\etal (1991). In order to compare the \PZ and cluster model fields we
smooth both with a Gaussian filter of width 20 \hmpc. Such a large smoothing is
required because of the large intercluster separation.

Fig.~15 shows the smoothed density fields along the Supergalactic plane,
within a distance of 150\hmpc, derived from the \PZ (upper panel) and
cluster (lower panel) catalogues. Overdensity contours are plotted in
steps of $\Delta \delta=0.2$ for the galaxy field, and of $\Delta
\delta=0.88$ for the cluster field. This is equivalent to rescaling the
cluster field by a factor $b_c=4.4$ which, as we shall see in \S5.2 below,
is our estimate for the relative linear bias parameter between IRAS galaxies and galaxy
clusters. As before, the heavy line traces the $\delta =0$ contour. The
dashed lines show the approximate location of the zone-of-avoidance in the
\PZ ($|b|\le 8^{\circ}$) and cluster ($|b|\le 20^{\circ}$) samples.

Despite the $b_c=4.4$ rescaling, the density peaks appear more prominent
in the cluster sample than in the \PZ map. This may reflect more severe
undersampling of the cores of rich clusters by IRAS galaxies than was
indicated by Baker \etal (1998), or it may be due to shot noise in the
cluster sample which, in spite of the heavy smoothing, is still
substantial. It should be noted, however, that the Great Attractor
region ${\rm (SGX,SGY)}=(-50,0)$ and the Perseus--Pisces supercluster
${\rm (SGX,SGY)}=(50,-20)$, where the effect is strongest, both lie within
the zone-of-avoidance of the cluster sample. It is therefore possible that
the height of these peaks has been artificially amplified by the coupling
between shot noise and the filling--in procedure. For this reason we
shall exclude the region at $|b|\le 20^{\circ}$ from the quantitative
analysis in \S5.2 below. Above $|b|= 20^{\circ}$ the Shapley
Concentration ${\rm (SGX,SGY)}=(-120,70)$ is the only peak with a larger
amplitude in the cluster map than in the \PZ map.
Amplitudes aside, the positions of the peaks in the two density maps are
very similar. The low density regions are also approximately coincident,
although they are less extended in the cluster map in which voids also
appear somewhat shallower. The latter effect simply reflects the fact that
the assumption of local biasing, $\delta_c=b_c\delta_g$, breaks down at
low $\delta_g$ because of the constraint $\delta_c\ge-1$. 

Similar
considerations apply outside the Supergalactic plane.
Fig~16 extends the comparison between clusters (left-hand side) and \PZ
galaxies (right-hand side) to four slices parallel to the Supergalactic
plane, at SGZ=$\pm45$ \hmpc and SGZ=$\pm 100$ \hmpc. Although some
discrepancies in the position of peaks and voids exist, the large--scale
features are remarkably similar across all the slices. In particular, the
large overdense region at positive SGZ is evident in both the clusters and
\PZ density fields. Unlike in the Supergalactic plane, there appears to be
no appreciable systematic difference between the amplitudes of density
peaks of clusters and IRAS galaxies.

Fig.~17 shows the velocity fields inferred from the two samples, again with
a 20\hmpc Gaussian smoothing, in a slice along the Supergalactic plane.
The amplitude of the velocity vectors is on an arbitrary scale but, for
the clusters, the velocities have been rescaled by the factor $b_c=4.4$.
The two large overdensities lying within the zone-of-avoidance in the
lower panel of Fig.~15 largely determine the infall patterns at those
locations [${\rm (SGX,SGY)}=(-50,0)$] and [${\rm (SGX,SGY)}=(50,-20)$]. 
Beyond $|b|=20^{\circ}$, however, the galaxy and cluster velocity fields
are very similar. Both exhibit a large coherent flow along the Camelopardalis,
Great Attractor, Shapley Concentration baseline. The infall onto the
Shapley region is more prominent in the cluster map but, in general, the
velocity field patterns are very similar at positive SGY, with an outflow
at positive SGX, and a convergent flow towards the Shapley Concentration
at negative SGX. Below the zone-of-avoidance, at negative SGY, both maps
show outflow from the Sculptor void, but this is less prominent in the
clusters map.

Note that a similar good agreement would have been obtained by considering
the density--velocity maps by Scaramella (1995) which also use the Abell/ACO
cluster distribution.
 
\subsection{The relative linear bias between galaxy clusters and 
IRAS galaxies.} 

If biasing is a local process then, in the regime where mass fluctuations
are small, we expect the amplitude of the bias to be independent of scale
(see e.g. Cole \etal 1998). This expectation is consistent with the
results of POTENT analyses performed by Dekel \etal (1993) and Sigad \etal
(1998) and we expect it also to be valid in the present analysis in which
the \PZ and cluster density and velocity fields have been smoothed with a
20\hmpc Gaussian filter. Thus, on the scales of interest, we expect the
two model density and velocity fields to be linearly related:
\begin{equation}
C=Pb_c+A_c,
\end{equation}
where $C$ and $P$ stand for cluster and \PZ and represent either
$\delta$ or any one of the Cartesian components of the velocity field.  The
constant $b_c$ is the bias parameter of clusters relative to IRAS galaxies
and $A_c$ allows for a relative offset in the mean density or bulk
velocities of the two fields. For quantitative analyses it is most
convenient to use the SGY Cartesian component of the velocity field
since this is the least affected by uncertainties in filling in the
zone-of-avoidance. To estimate $b_c$ we adopt the strategy of Hudson \etal
(1995) and Branchini \etal (1997), already used in \S4.2, of regressing
the two model fields by minimising the quantity
\begin{equation}
\chi^2=\sum_{i=1}^{N_t}{(C_i-A_c-b_c P_i)^2\over(\sigma_{C,i}^2+b_c^2
\sigma_{P,i}^2)},
\label{eq:chi0}
\end{equation}
where the subscript $i$ refers to any of the $N_t$ gridpoints within the
comparison volume. The quantities $\sigma_{C,i}$ and $\sigma_{P,i}$
represent the errors in the cluster and \PZ fields, respectively.

The errors in the cluster field, $\sigma_{C,i}$, have been estimated by
Branchini \etal (1997). They are the sum in quadrature of the intrinsic
errors in the reconstruction procedure, as estimated by BP96 using
Montecarlo techniques, and the shot noise uncertainties which are dominant
for the very sparse cluster fields. The latter were evaluated using the
mock catalogue generated by Kolatt \etal (1996) which is designed to
reproduce the distribution of structures in our local universe.  A typical
error in the cluster overdensity field is
$\langle\sigma_{C,\delta}\rangle=0.36$, and in the SGY--component of the
velocity field it is $\langle\sigma_{C,v_y}\rangle=250 \beta$ \kms.

The errors in the \PZ fields have been estimated using the mock catalogues
described in \S3.2. The basic procedure consists of comparing the density
and velocity fields obtained by applying Method~1 to the  mock galaxy
catalogues with the true fields in the parent N--body simulation. Fig.~18
(which is analogous to Fig.~4) compares the true and reconstructed density
and velocity fields of a mock \PZ catalogue (different from the one shown
in Fig.~4) in a slice along the Supergalactic plane (upper panels). The
maps of the density  and velocity errors are shown in the lower panels.
Both fields are smoothed with a Gaussian filter of radius 20 \hmpc. As in
Fig.~4, the errors in the density and velocity reconstruction occur
preferentially within the zone-of-avoidance. However, the correlation with
the signal, which was present in the outer regions of  Fig.~4, has been
almost erased by the large smoothing. We have characterized the reconstruction
errors by noticing that the residuals correlate with distance, galactic
latitude and, for the velocity field, with the signal itself. From the
analysis of the mock catalogues, we have derived two approximate
expressions for the errors (valid for $|b|>20^{\circ})$:
\begin{equation}
\langle\sigma_{P,\delta}\rangle = 0.11- 6.7 \ 10^{-4} \cdot |b| + 3.3 \
10^{-4} \cdot r
\label{eq:sd}
\end{equation}
and
\begin{equation}
\langle\sigma_{P,v_y}\rangle = 70+ 2.7 \ 10^{-1}\cdot r+ 0.15 \cdot |v_y|,
\label{eq:sv}
\end{equation}
where $r$ is the distance measured in \hmpc , $b$ is the galactic latitude
in degrees (not the linear biasing factor), 
and $|v_y|$ is the amplitude of the SGY Cartesian component
in \kms. The actual assumed error at a gridpoint has been generated from a
Gaussian distribution centred on $\langle\sigma_{P}\rangle$ with
dispersions of 0.04 for the overdensity and 40 \kms for the $v_y$ field,
respectively.

The regression of the \PZ and cluster fields uses all the gridpoints
within 120 \hmpc and outside the cluster zone-of-avoidance (i.e. at
$|b|>20^{\circ}$). 
As a result of the large smoothing applied, not all the gridpoints in the
comparison volume are independent. The number of independent points, $N_i$,
can be computed as in Dekel \etal (1993), 
\begin{equation}
N_i^{-1}=N_t^{-2}\sum_{j=1}^{N_t}\sum_{i=1}^{N_t} \exp(-r_{ij}^2/2R_s^2),
\label{eq:ind}
\end{equation}
where $r_{ij}$ is the separation between gridpoints $i$ and $j$ and $R_s$
is the smoothing radius of the Gaussian filter. This expression has been
derived for a smooth density field and its application to a velocity field 
requires some caveats. For a given comparison volume, the large
coherence of the velocity field causes the number of independent points to
be smaller than $N_i$. On the other hand, our peculiar velocities are
predicted using a sample covering a larger volume than that used in the
comparison and this increases the number of independent points. Because of
these competing effects, we simply approximate the number of independent
points by $N_i$, even in the $v-v$ comparison.

As in \S4.2 we define the $\chi^2_{eff} \equiv (N_i/N_t) \chi^2$ statistic
which corresponds, in practice, to multiplying the errors $\sigma_{P}$ and
$\sigma_{C}$ by $\sqrt{N_t/N_i}$ in equation~(\ref{eq:chi0}).  We use this
statistic to estimate the errors in $b_c$ and $A_c$.

Fig.~19 (upper panel) 
shows a $\delta$--$\delta$ scatterplot of the model cluster and 
\PZ overdensity fields measured at $\sim 1000$
randomly selected gridpoints. All 6426 original points are used in the regression
analysis.
The solid line shows the best fit obtained by minimizing $\chi^2_{eff}$ and
the parameters of the fit are listed in Table~3. The resulting bias
parameter is $b_c^{\delta}=4.4\pm0.6$. This is consistent with results 
from an independent likelihood analysis in which the phenomenological
power spectra of IRAS galaxies and Abell clusters were compared
(e.g. Peacock and Dodds 1994).
The parameter $S^{\delta}=\chi^2_{eff}/N_{dof}=1.14$, 
which may be taken as an
indication that the errors have not been grossly over- or underestimated.
The systematic difference in the amplitude of the density peaks in
the cluster and IRAS $\delta$ fields, noticed in Fig~15, manifests itself
as a deviation from the best fitting line at large $\delta_{PSCz}$.
Restricting the regression to values of $\delta_{PSCz}\le 0.35$ 
returns $b_c^{\delta}=4.5\pm0.6$, almost
identical to the previous value, and  
$S^{\delta}=\chi^2_{eff}/N_{dof}=1.15$. This suggests that the exact 
weighting of \PZ
galaxies in high density regions has only a minor effect on the regression
analysis, mainly because the overdensity mismatch in large density peaks is
restricted to a very few gridpoints. As a further check, we have repeated the
$\delta$--$\delta$ regression using a \PZ velocity
model derived without applying our standard procedures for collapsing
clusters and handling triple valued regions (which should exacerbate any
discrepancies associated with high peaks.)  The results, listed in the
second row of Table~3, show that the effect on $b_c$ is indeed very small,
leading to $b_c^{\delta}=4.5\pm0.6$.

The lower panel of Fig.~19  shows the scatterplot of the SGY Cartesian
components of the two velocity fields in which, as in the upper panel, 
to avoid overcrowding, we only show $\sim 1000$ randomly selected
gridpoints. As shown in Table~3 the slope of the best fitting line is 
$b_c^{v_y}=4.7\pm0.6$ with no significant zero point offset,
indicating that the SGY components of the cluster and \PZ bulk flows are
consistent with one another over the scales of interest. However, as
indicated in Table~3, the resulting $\chi^2_{eff}$ is large,
$S^{v_y}=1.55$. 
This could be due to an underestimate of the errors since the error
analysis by Branchini \etal (1997) is based on clusters from the Kolatt \etal 
(1996) mock catalogues and these do not accurately match the Abell/ACO cluster
distribution and velocities. 
We can obtain $S^{v_y}\simeq 1.0$ if we allow for a reasonable error 
underestimate in the cluster field of $\sim 30 \%$, in which case we 
obtain $b_c^{v_y}=4.0\pm0.6$. Note that consistent values of
$b_c^{\delta}$ and $b_c^{v_y}$ are obtained within 1-$\sigma$ whether or
not the cluster errors have been underestimated at this level.

It is worth emphasizing that the agreement of the $\delta$--$\delta$ and
$v_y$--$v_y$ comparisons is not trivial. The $\delta$-$\delta$ comparison
is local; it is hardly affected by problems related to filling in masked
regions but is potentially prone to the cluster core weighting problem.
The $v_y$-$v_y$ comparison, on the other hand, involves the distribution of
objects within the entire sample and thus is much more strongly affected by
the unknown mass distribution within the zone-of-avoidance and beyond the
sample's edge. We might therefore expect the two comparisons to be subject
to different biases.  The  agreement in the estimates of $b_c$ from the two
analyses suggests that systematic biases have been properly taken into
account and that the linear biasing assumption is a good approximation, at
least on scales larger than our 20 \hmpc smoothing.

\section{The bulk velocity vector} 
\label{sec:vlb}

In this section we consider the peculiar velocity, $\vv_b(R)$, of spheres
of radius $R$ centred on the Local Group. This is a low order statistic
that, in principle, can be estimated observationally. 
The expectation value of the bulk velocity, $|\vv_b|$,
averaged over scale $R$ is:
\begin{equation} 
\langle |\vv_b(R)|^2 \rangle = {{\beta^{2}H_0^2}\over{2\pi^2}} \int
P(k) W(kR)^2 dk,
\label{eq:bv}
\end{equation}
where $P(k)$ is the power spectrum of density fluctuations, 
$W(kR)=3{{\sin(kR)-kR\cos(kR)}\over{(kR)^3}}$ is the Fourier transform of
the spherical top hat window function in real space, chosen to facilitate 
comparison between the theoretical
definition (eqn.~\ref{eq:bv}) and its observational analogue 
(eqn.~\ref{eq:bv2}).
It is worth stressing that eqn.(~\ref{eq:bv}) assumes equal volume
weighting while the use of different weighting schemes
may lead to  quite different results
(Kaiser 1988,  Strauss et al. 1995,  Giovanelli et al. 1998). 
If the initial fluctuation field obeys Gaussian statistics, then evolution through
gravitational instability preserves a Gaussian distribution for the
amplitude of each Cartesian component of $\vv_b(R)$, so that $|\vv_b(R)|$
has a Maxwellian distribution. This property makes it difficult to
constrain  $P(k)$ from the measured  $\vv_b(R)$. Nevertheless, comparison
of the measured $\vv_b(R)$ with the velocities predicted from the \PZ
gravity field allows, in principle, an estimate of the $\beta$ parameter.
In practice, however, the bulk velocity is extremely sensitive to
systematic errors both in the observational data and in the models. We
attempt to take this carefully into account in the following analysis.

\subsection{The model bulk flow} 
\label{sec:vlbm}

The basis for our treatment of random and systematic errors in the model
bulk velocity vector are, again, the mock \PZ catalogues described in
\S3.2.  Finite volume effects are not a concern here since the volume of
the simulation box is comparable to that of the \PZ sample.  However,
neglecting modes on scales larger than the size of the sample can bias the
comparison between the model bulk flow and the measured one. In what
follows we will treat these two effects separately. We first account for
the intrinsic errors in the bulk flow model and we then address the problem
of minimizing the effect of missing large-scale power.

From each mock catalogue we generate a model velocity field using Methods~1
and~2 of \S3.2. The field is then smoothed onto a 64$^3$ cubic grid of
side 192 \hmpc using a 12\hmpc Gaussian filter. The 
smoothing filter used here is different from the one used before because
our aim is to perform a homogeneous 
comparison with the Mark III and SFI bulk flows. 
We measure the cumulative
bulk velocity vector in the CMB frame by averaging over the peculiar
velocities measured at gridpoints:
\begin{equation} 
\vv_b(R) = {{\sum_{(i,j,k)<R}\vv_{i,j,k}}\over{\sum_{(i,j,k)<R}}},
\label{eq:bv2}
\end{equation}
where $\vv_{i,j,k}$ is the predicted velocity vector in the CMB frame at
gridpoint $(i,j,k)$. The sum $\sum_{(i,j,k)<R}$ extends over all the
gridpoints contained within a sphere of radius $R$. The same exercise is
repeated using the  original N--body velocity field. This gives an
unbiased estimate of the true bulk velocity and the 
comparison between true and reconstructed velocities is an 
estimate of the error in the \PZ model bulk flow. 

In the upper panel of Fig.~20 we show the
average difference between reconstructed and true  
cumulative bulk velocity vectors.
Each Cartesian component is displayed 
with a different symbol. The errorbars, representing the dispersion around
the mean, are shown only for the SGY component. 
The dispersions for the remaining two components are larger by nearly a
factor of two. From the plot we see that the errors in the SGY and SGZ 
components of the reconstructed bulk flow are mainly random errors, but the
SGX component is affected by a systematic positive bias. 
The amplitude of this  systematic error decreases with distance
(from  $\sim 61$ \kms at 10 \hmpc to $\sim 16$ \kms at 100 \hmpc). 
Very similar results are
obtained if the reconstruction is carried out using Method~2.
These systematic errors arise mainly from the filling--in procedure
for the zone-of-avoidance which, as we discussed in \S 3.2, affects 
both the velocity fields  and, via uncertainties
in modeling the LG velocity, 
the transformation from LG to CMB frame. To corroborate this hypothesis
we have performed 10 reconstructions using unmasked mock catalogues. 
With full sky coverage, the
systematic bias in the SGX component disappears and
the dispersions in the SGX and SGZ components decrease by 
a factor of two.
The systematic error in the SGX component is ultimately induced by the
requirement of having only LG--like observers in the \PZ mock catalogues. 
In particular, visual inspection reveals that 
the galaxy dipole constraint in \S 3.2 almost invariably implies the
existence of a Great Attractor--like density peak at a distance 
of 30--60 \hmpc, lying at negative SGX and partially
overlapping the zone-of-avoidance. The filling--in procedure 
seems to underestimate the amplitude of this partially hidden density peak
and this in turn causes a systematic offset in the SGX component 
of the reconstructed peculiar velocities, with minor effects on the two
other components, as shown in Fig.~20. 
A similar bias is to be expected in the true model \PZ bulk flow
because the Great Attractor really exists in our universe.
To obtain bias-free estimates of the \PZ bulk flows, we must correct 
for such systematic errors. Clearly, the best way of circumventing the problem
is by restricting attention to the SGY component of the bulk
velocity. However, sometimes one is interested in the 
amplitude of the bulk velocity and, in this case, 
a direct numerical correction of the systematic error is appropriate. 

Since the amplitude of the estimated errors depends on $\beta$, the errors
displayed in the upper panel of Fig.~20 (or the analogous quantity for the
velocity amplitude) cannot be directly used to make quantitative
corrections to the \PZ model bulk flow.  However, we can take advantage of
the fact that, to first approximation, the model velocities scale linearly
with $\beta$. Thus, the ratio between reconstructed and true bulk
velocities should be independent of $\beta$. This ratio, averaged over our
mock catalogues, is shown by filled circles in the lower panel of Fig.~20,
as a function of distance, with errorbars again representing the dispersion
around the mean. Since it is independent of $\beta$, this ratio may be used
as a multiplicative factor to correct the predicted \PZ bulk flow
amplitudes for systematic errors. As expected, the corrections become very
small when considering the SGY component (open circles). We have performed
a similar analysis to estimate errors in the \PZ bulk velocity obtained
using Method~2 and in the 1.2Jy bulk flow model, obtained using Method~1.

We are now ready to derive model bulk flows and their
uncertainties from the \PZ and 1.2Jy surveys. The upper left panel in 
Fig~21 shows the amplitude of the cumulative bulk velocity
vector, predicted using various samples and methods. 
In all cases the bulk flow was computed by integrating the density
field out to the same distance of 200 \hmpc, hence enforcing 
the same volume bias in all cases. The
velocities are normalized to $\beta=1$. Open and filled circles show
\PZ results using Methods~1 and~2 respectively, and filled squares show
results from the 1.2Jy survey, all corrected for systematic errors as
discussed in the preceding paragraph. For clarity only the 1--$\sigma$
error bars from Method~1 are plotted. The filled triangles show the bulk
velocity computed from the Abell/ACO model velocity field, rescaled by
$b_c=4.4$ (see \S5.2). There is remarkably good agreement between the
cumulative bulk flows computed from the different sets of mass tracers,
with typical deviations of less than $\sim 20\%$ from the mean in the
amplitude from the different surveys and analysis methods. The directions
of the cumulative bulk flow vectors, measured within 60 \hmpc, are plotted
in the upper right-hand panel of Fig.~21.
The \PZ model bulk flow points, within
1$\sigma$, to the direction of both the 1.2Jy and Abell/ACO model bulk
flows. The direction of the CMB dipole, indicated by the
asterisk, is plotted for reference at the centre of the  figure.
In the lower panel of Fig.~21 we show the misalignment angle of the various
bulk flows with respect to the \PZ--Method~1 model.
In all cases the misalignment is small and almost independent of 
radius. The only exception is the increasing misalignment angle at large 
radii between the \PZ bulk velocities determined using Method~1 and Method~2.
Note the close alignment between the \PZ and clusters bulk
flows at all radii.

So far we have modelled bulk velocities in the nearby universe by
considering only the gravitational effect of the mass distribution within 200\hmpc.
Bulk flows, however, are
sensitive to the mass distribution on scales larger than those probed
by our samples. Neglecting large scale modes leads to a 
volume bias which can affect the comparison with observations. 
Various methods have been proposed to restore the missing large scale 
modes (e.g. Strauss \etal 1995, Tormen and Bertschinger 1996, Cole 1997).  
Here we propose a simple statistical treatment based 
on linear theory, using eqn.~(\ref{eq:bv})
to obtain a correction factor for the missing contribution
to the amplitude of the bulk velocity from scales 
beyond the sample boundary. 
The correction factor is the ratio
\begin{equation} 
F(P(k),k_{min})= {{\int_{0}^{\infty} P(k) W(kR)^2 dk}\over{
\int_{k_{min}}^{\infty} P(k) W(kR)^2 dk} }. 
\label{eq:bvratio}
\end{equation}
The numerator in this equation is the mean true bulk flow, while the
denominator is the mean bulk flow generated only by density fluctuations
on scales smaller than $R_{max}={{2 \pi} \over {k_{min}}}$. In our case,
$R_{max} \sim 200$\hmpc. Note that the ratio depends only on the spectral
shape and on ${k_{min}}$; the dependence on $\beta$ cancels 
out. Tests using large N--body simulations (e.g. Jenkins \etal 
1998) have shown that this approach is indeed effective.
Since the amplitude of the bulk flow follows a Maxwellian distribution, and 
its components are Gaussian random variables, we can numerically
estimate the uncertainty in this correction.
Assuming that these uncertainties and the intrinsic random errors
computed above are uncorrelated, we obtain the total random uncertainties on 
the model bulk flow by adding them in quadrature.
In computing $F$ we assume a CDM model with spectral shape
$\Gamma=0.25$, as suggested by a wide variety of data on the large-scale
galaxy distribution (e.g. Baugh 1996).
%

We have applied the correction of eqn.~(22) to the bulk velocity
reconstructed from the \PZ survey using Method~1. The resulting cumulative
bulk velocity (for $\beta=1$) is shown in Fig.~21 as a dot--dashed line
labelled M1L.  We regard this as our best estimate of the
$\beta$--dependent bulk flow.


\subsection{Model vs. observed bulk flow} 
\label{sec:vlbo}

In this Section, we compare our best estimate of the predicted bulk flow
velocity from the \PZ survey with recent observational estimates. This 
comparison serves two purposes. Firstly, consistency between predicted
and observed velocities lends support to the hypothesis that structure
grew by gravitational instability and gives confidence in the integrity of
the observational data. Secondly, the comparison allows an estimate of the
parameter $\beta=\Omega_0^{0.6}/b$.

The determination of bulk flows from peculiar velocity surveys is prone to
systematic errors. For example, zero-point errors in the calibration of
the distance indicators coupled with limited sky coverage may mimic a bulk
flow. Similarly, the coupling of large intrinsic errors in distance
measurements with inhomogeneities in the density distribution (the
inhomogeneous Malmquist bias) also results in a spurious outflow. Thus, 
to perform an unbiased comparison with theoretical predictions requires
full-sky, homogeneous surveys of peculiar velocities and an accurate model
of the survey's window function. 

In Fig.~22 we plot observational determinations of bulk flows derived from
four (almost) independent datasets. The lower panel shows the full cumulative
bulk flow as a function of distance while the upper panel shows its SGY
component only. The filled triangles (taken from Dekel \etal 1998)
represent the cumulative bulk flows in the Mark~III catalogue.  The sparse
and noisy Mark~III velocities have been smoothed assuming that the velocity
field is irrotational. This guarantees that a unique three--dimensional
velocity field is derivable from the observed radial velocities, as in the
POTENT method (Dekel, Bertschinger \& Faber 1990). The resulting
three--dimensional peculiar velocity field, smoothed with a 12\hmpc
Gaussian, and defined on a regular grid is directly comparable to our model
prediction. The filled square shows the preliminary result by Eldar \etal
(1998) who use a POTENT smoothing technique to derive the bulk velocity
from the SFI catalogue (Giovanelli \etal 1997a, 1997b).  The open triangle
in the lower panel is the bulk velocity inferred from 44 supernovae Type~Ia
by Riess, Press \& Kirshner (1995), as reported by Dekel (1997) (the
effective radius of this last measurement is much smaller than the depth of
the sample because the data were weighted by the inverse of the
errors). Finally, the open square at large distance shows the bulk velocity
derived by Lauer \& Postman (1994; LP94) from a sample of brightest cluster
galaxies. The directions of the observed bulk velocity vectors are given in
the right-hand panel.  For the Mark~III, SFI and Sn 1a determinations, they
have been estimated from data within 50 \hmpc.  The direction of the LP94
bulk flow is taken from Strauss (1997) and refers to a depth of $\sim 90
\hmpc$.

Except for the LP94 result, there is excellent agreement between the
various determinations of the bulk flow in Fig.~22. These may be compared
with the bulk flows predicted by the \PZ survey, indicated by the thick
dot--dashed lines which enclose the 1--$\sigma$ allowed range. These are
normalized to $\beta=0.75$ and corrected for volume bias.  The filled
circle in the right-hand plot marks the direction of the predicted bulk
velocity vector within 50 \hmpc.

Requiring that the predicted bulk flow should match the
measured one gives an estimate of $\beta$.  In performing this comparison
we shall ignore the discrepant LP94 result which Strauss \etal (1995) and
Watkins and Feldman (1995), amongst others, have argued is inconsistent
with currently acceptable cosmological models.  The LP94 data point is also
inconsistent with our predicted bulk flow derived only from the
gravitational instability and linear biasing hypotheses. Indeed, as shown
in Fig.~22, matching the amplitude of the LP94 bulk flow would require a
value of $\beta \simeq1.9$, which is incompatible with all other current
measurements (e.g. Giovanelli \etal 1998a, 1998b, Dekel \etal 1998 and
Eldar \etal 1998) and implies a very large misalignment angle of
$70^{\circ}$ between the LP94 vector and the direction of our model flow
velocity.  In what follows we limit our comparison of model and observed
bulk flows by imposing a series of restrictions designed to minimize
possible systematic errors.  These restrictions are:

\noindent
-- We consider only the cumulative bulk flow from Mark~III as estimated by
Dekel \etal (1998). Like the \PZ model velocity field, the Mark~III velocities
have been smoothed onto a regular grid and filtered on a similar scale.
This ensures that the comparison is as homogeneous as possible.

\noindent
-- We consider Mark~III peculiar velocities calibrated according to 
the recent VELMOD analysis by Willick and Strauss (1998).

\noindent
-- We exclude scales smaller than 30\hmpc since small differences in the smoothing
procedures applied to model and observed velocities in the nearby
region can bias the comparison.

\noindent
-- We do not include the result of Eldar \etal (1998) which, although
consistent with those of Dekel \etal (1998), is still preliminary.

We perform the comparison between model and observed bulk flows using a
likelihood technique similar to that used by Strauss \etal (1992a) and
Schmoldt \etal (1998).  The aim is to estimate the likelihood of a
particular value of $\beta$ given the bulk flow observed in the Mark~III
catalogue and the one derived from the \PZ survey using Method~1.  Under
the assumption that the bulk velocities are Gaussian random fields it is
easy to show that the joint probability distribution for the observed and
model bulk flows is a multivariate Gaussian that depends only on its
covariance matrix. The latter is completely specified by the value of
$\beta$ and by the power spectrum of density fluctuations, $P(k)$.  In this
work we restrict attention to the family of CDM models for which $P(k)$,
parametrised as in Davis \etal (1985), is completely specified by its shape
parameter, $\Gamma=\Omega_0 h$, and normalization, $\sigma_8$.  Although the latter can
be regarded as an independent parameter, adopting the normalization
inferred by Eke, Cole, and Frenk (1996) from the observed abundance of
galaxy clusters, allows one to relate $\sigma_8$ to $\Omega_0$.  In the
following we consider only two cases, $\sigma_8$ = 0.52 and 0.87, which
correspond to the cases of a critical, $\Omega_0=1.0$, and an open, 
$\Omega_0=0.3$, universe. (For a flat Universe with $\Omega_0=0.3$, the
required value of $\sigma_8$ is similar).

Given a cosmological model specified by $\sigma_8$, $\beta$, and $\Gamma$,
we compute the joint probability of obtaining the cumulative bulk flow
measured by Dekel \etal (1998) at 30, 40, 50 and 60 \hmpc (${\bf
v}_{M}(i), i=1,4$), and the \PZ--M1 cumulative bulk flow displayed in  Fig.~22, at all
radii from 10 to 80 \hmpc $({\bf v}_{P}(j), j=1,8)$. We ignore the
correction for volume bias. The resulting probability density is: 
\begin{equation}
f({\bf v}_M(i),i = 1,4, {\bf v}_{P}(j), j=1,8)=
(2\pi)^{-3(4+8)/2} (det M)^{-3/2} \exp(-\frac{1}{2} {\bf v}_l \cdot
{\bf v}_m (M^{-1})_{lm}) d\bv_M d\bv_P,
\label{fv}
\end{equation}
where the covariance matrix $M_{lm}$ is: 
\begin{equation}
M_{lm} 
= \frac{1}{3} < {\bf v}_l \cdot {\bf v}_m > \nonumber
= \frac{H_{\circ}^{2} \beta^{2}}{6 \pi^2} 
\int dk P(k) \tilde{W}_l(k) \tilde{W}_m(k). 
\label{covariancematrix}
\end{equation}
The window functions are:
\begin{equation}
\tilde{W}_M(k) = 3 \frac{\sin(kr_i) - kr_i
\cos(kr_i)}{(kr_i)^3}
\end{equation}
for the Mark~III cumulative bulk flow measured within $r_i$ and
\begin{equation}
\tilde{W}_P(k) = 3 \frac{\sin(kr_j) - kr_j\cos(kr_j)}{(kr_j)^3}
- 3 \frac{\sin(kR_s) - kR_s\cos(kR_s)}{(kR_s)^3}
\end{equation}
for the \PZ cumulative bulk flow measured within $r_j$. In eqn.~(26),
$R_s=200$ \hmpc represents the outer radius of the \PZ sample, i.e. the
maximum distance out to which the \PZ density field is integrated to
predict peculiar velocities.  The second term in the right-hand side of
eqn.~(26) accounts for the volume bias affecting the M1--\PZ bulk flow. 
This correction is robust in the sense that the alternative strategy of
limiting the integration in eqn.~(\ref{covariancematrix}) to $k\ge k_{min}
=2\pi/R_s$ gives very similar results.  Finally, we take into account random
errors in the measured and modeled bulk flows by adding them in quadrature
to the diagonal terms of the covariance matrix.

From the joint probability density distribution (eq.~\ref{fv})
we construct the relative likelihood,  ${\cal L}$, of different 
world models $(\beta, \Gamma, \sigma_8)$:
\begin{equation}
{\cal L} = -2 \ln(f).
\label{lf}
\end{equation}
The two plots in Fig.~23 show likelihood contours of ${\cal L}({\bf
v}_M(i),i = 1,4; {\bf v}_{P}(j), j=1,8)$  in the $(\beta,l_{eq}=1/\Gamma)$
plane, plotted at confidence levels of 68, 90 and 99 \%. The shape of these
contours is very similar for the two normalizations although the
range of allowed values of $\beta$ is somewhat larger for the high
normalization.  In both cases the likelihood peaks at $\beta=0.6-0.7$; low
values of $\beta$ are excluded at high confidence level primarily by the
requirement to match the amplitude of the observed bulk flow.  No
stringent constraint can be imposed on the shape parameter $\Gamma$ apart
from an upper limit that follows from the fact that the power cannot be
confined to very small scales. The cluster normalization also defines a
$(\beta,\Gamma) \rightarrow (b,h)$ transformation. Therefore, we can
explore the effect of introducing an observational constraint on the Hubble
parameter, ($0.4<h<0.8$), which defines a vertical region in the
$(\beta,\Gamma)$ plane. Introducing this extra constraint has a major
effect in the case of low normalization since it forces the power to
originate from unacceptably small scales. By contrast, in the case of
high normalization, the extra constraint on $h$ is perfectly consistent
with the likelihood contours.

Following Schmoldt \etal (1998) we can set a more stringent constraint on
$\beta$ by considering not only the bulk flow but the entire set of
observational information available on peculiar velocities in the local
universe.  In particular, we use the CMB dipole (${\bf v}_{LG}$), the
amplitude of the local velocity shear ($|{\bf v}_s|$), and the dipole in the 
gravitational acceleration induced by the distribution of \PZ galaxies
in $N$ independent radial shells (${\bf v}_d(k), k=1,N$) in order to
compute the constrained probability function $f({\bf v}_M(i), i=1,4; {\bf
v}_{P}(j), j=1,8; {\bf v}_d(k), k=1,N; {\bf v}_{LG} | {\bf v}_s$). The
modeling of this probability distribution, its covariance matrix and the
treatment of the errors may be found in Schmoldt \etal (1998) and Strauss \etal
(1992a) to which we refer the reader for a more detailed discussion.  In
this analysis we use the \PZ differential dipole, ${\bf v}_d$, measured by
Schmoldt \etal (1998) in $N= 15$ non-overlapping top-hat windows of width
$10$ \hmpc and the same values of ${\bf v}_{LG}=625\pm25$ and $|{\bf
v}_s|<200$ \kms used in \S 3.2.  

The effect of adding these new observational constraints is shown in
Fig.~24. The overall shape of the likelihood contours does not change
appreciably. However, the valley along the $\Gamma$ direction is now deeper
and so the value of $\beta$ is better constrained.  This is clearly seen in
Fig.~25 where we plot the distribution of the normalized likelihood of
$\beta$ obtained by marginalising the joint likelihood distribution over
$\Gamma$. The continuous lines give the marginal distributions after
integrating over all values of $\Gamma$, while the dot-dashed lines give
the result of limiting $\Gamma$ to the values allowed by the extra
constraints on $h$.  For $\sigma_8=0.87$, we find $\beta=
0.6^{+0.22}_{-0.15}$ (1--$\sigma$), irrespective of the constraint on
$h$. Thus, in a low density universe, a power spectrum normalized according
to the cluster abundance is consistent with the velocity data.  By
contrast, for the low normalisation, $\sigma_8=0.52$, the constraints on
$\beta$ change appreciably with the addition of the new constraints,
leading to a larger value of $\beta= 0.9^{+0.35}_{-0.20}$ (1--$\sigma$).
As may be seen in Fig.~24, however, in this latter case the relative
probability within the strip allowed by the constraint on $h$ is smaller
than in the case of high normalization, thus rendering a high $\Omega_0$
universe less likely according to the velocity data.  Finally, it is
interesting to note that the assumption of linear bias coupled with the
cluster normalization of the power spectrum implies that $\beta \simeq
0.5/{\sigma_8^{PSCz}}$, irrespective of the value of $\Omega_0$. From the
variance of the \PZ galaxy counts at 8~\hmpc, $\sigma_8^{PSCz}\simeq
0.7$, and so we obtain $\beta \simeq0.7$, consistent with the estimates
above at the 1--$\sigma$ level.

\section{Discussion and conclusions}
\label{sec:conc}

The recently completed \PZ  redshift survey of IRAS galaxies represents
an almost ideal dataset for studying the mass distribution and the gravity
field in the local universe. In this paper we have used the \PZ survey to
develop a nonparametric model of the local cosmic velocity field which
supersedes results derived from the shallower 1.2 Jy and the sparser
QDOT surveys. Our reconstructions are based on the assumptions that cosmic
structure has grown by gravitational instability and that fluctuations in
the galaxy distribution are proportional to fluctuations in the
mass distribution. We have paid particular attention to a careful
estimation of random and systematic errors using a suite of mock \PZ
catalogues constructed from large cosmological N--body simulations. As a
further check on the validity of our results, we have implemented two
independent methods for reconstructing the \PZ model velocity fields, 
both of which give consistent results.

Because of the large size of the \PZ survey, the density and velocity
fields can be reliably reconstructed out to a depth of
150\hmpc. We have presented maps of the galaxy distribution which, with a
smoothing of XS 6\hmpc, clearly show the relative sizes of the main
structures that characterize our local universe. The two largest peaks in
our neighbourhood are the Great Attractor, made up of the
Hydra--Centaurus and Pavo--Indus--Telescopium superclusters, and the
Perseus--Pisces supercluster located in roughly opposite directions along 
the Supergalactic plane.  The Local, Coma-A1367, and Camelopardalis
superclusters as well as the Cetus Wall are clearly visible in our
maps, as is the giant Shapley concentration which appears near the edge of
the survey, behind the Great Attractor. The largest underdensity in our
vicinity is the well-known Sculptor void, but this is almost matched in
size by a void in the background of the Camelopardalis supercluster. Three
more underdense regions that exert an important influence on the local
dynamics are the voids in the foreground of Coma, in the background of the
Perseus--Pisces complex, and behind the Great Attractor.

The local velocity field implied by the \PZ density field is complex. The
dominant features are the organized infall patterns towards the large mass
concentrations in the Great Attractor, Perseus--Pisces and Coma
regions. Superimposed upon these are impressive coherent flows along a
ridge between Cetus and Perseus--Pisces and along the
Camelopardalis--Virgo--Great Attractor--Shapley direction. We see no
prominent back-infall onto the Great Attractor. Instead, the flow in this
region is a result of an interplay between the compressional push of two
nearby voids and the pull of the Shapley concentration on much larger
scales.

The \PZ reconstruction of the velocity field agrees well with results from
the 1.2Jy survey within 80\hmpc, the region in which the latter provides
adequate sampling. The only noticeable difference are the predicted bulk
velocity vectors which (in the CMB frame) differ in the two surveys by
$\sim 130 \beta$ \kms in each Cartesian component. This discrepancy arises
from uncertainties in the way in which the zone-of-avoidance is filled in
and from the sampling noise of the density field at large distances.  While
the former affects the \PZ and 1.2Jy results equally, the latter is more
severe in the 1.2Jy case because of the larger shot noise. Thus, the
misalignment between the Local Group acceleration and the CMB dipole
vectors is $\sim 25^{\circ}$ for the 1.2Jy survey, but only $\sim
15^{\circ}$ for the \PZ survey (Schmoldt \etal 1998).  Comparison of the
spherical harmonic multipoles of the two model velocity fields in radial
shells confirms the consistency of the \PZ and 1.2Jy survey predictions in
the region of overlap (Teodoro \etal 1998).

As was the case with the 1.2Jy survey, the velocity field reconstructed
from the \PZ survey is partially inconsistent with the peculiar velocities
in the Mark~III catalogue. A visual inspection of the \PZ and 1.2 Jy
velocity maps (Figs.~6 and 9) shows some differences around the
Hydra--Centaurus complex, where the Mark~III data exhibit a large shear,
and in the Perseus Pisces region, which, in the Mark~III catalogue, appears
to participate in a large streaming motion (Davis, Nusser and Willick
1996).  As pointed out by these authors, beyond 30 \hmpc there is a large
coherent residual dipole in the difference between the 1.2Jy gravity field
and the Mark~III peculiar velocity field. This cannot be ascribed to
improper modelling of the gravity field by IRAS galaxies since the dipole
depends only on the mass within the surveyed volume. Furthermore, Baker
\etal (1998) have shown that the gravity field in the optical ORS survey
(Santiago \etal 1995, 1996) is consistent with the 1.2Jy gravity field (and
thus, by implication, with the \PZ field as well). This suggests that the
undersampling of cluster cores by the predominantly spiral galaxy
population in IRAS surveys does not have a large effect in the inferred
gravity field. The most plausible explanation for the discrepancy between
the Mark~III velocities and the IRAS (and ORS) model predictions seems to
be some systematic bias that affects the Mark~III catalogue beyond 30 \hmpc
(Baker \etal 1998, Willick and Strauss 1998).

As we saw in \S5.1, the density and velocity fields inferred from the
\PZ survey agree well with those inferred from a sample of Abell/ACO
clusters. This is perhaps surprising since clusters are selected in a very
different way from galaxies, but it is reassuring and suggests that
systematic errors are under control in both cases. Comparison of the two
model density and velocity fields, smoothed on the same cubic grid, out to a
distance of 140 \hmpc, gives the relative linear bias, $b_c$, between the
rich cluster population and \PZ galaxies. A simple $\chi^2$ analysis of
the density--density and velocity--velocity comparisons gives very similar
results, $b_c=4.4 \pm 0.6$ and $b_c=4.7 \pm 0.6$,
respectively. These values are consistent with the estimate, $b_c=4.5$,
obtained by Peacock and Dodds (1994) from an independent likelihood  analysis 
of phenomenological power spectra. 
A more detailed comparison of the relative bias field
of clusters and galaxies will be presented by Plionis \etal (1998).

Finally, averaging over the peculiar velocity field reconstructed from the
\PZ survey, we have calculated the expected bulk velocity of concentric
spheres around us.  Comparison of these predicted bulk flows with those
measured from the Mark~III catalogue by Dekel \etal (1998) gives an
estimate of $\beta=\Omega^{0.6}/b$.  We implemented a likelihood analysis
to carry out this comparison, taking into account the observed CMB dipole,
the observed local shear field and the velocity of the Local Group
predicted by the \PZ gravity field.  If a high normalization for the power
spectrum is assumed ($\sigma_8=0.87$), then our best estimate is $\beta=
0.6^{+0.22}_{-0.15}$ (1--$\sigma$).  This value of $\beta$ is consistent
with results from analyses of the \PZ dipole (Schmoldt \etal 1998,
Rowan--Robinson \etal 1998), and with the most recent determinations of
$\beta$ from velocity--velocity comparisons (Davis, Nusser \& Willick 1996,
Willick \etal 1997b, da Costa \etal 1998, Willick and Strauss 1998 and,
within 1-$\sigma$, also with Sigad \etal 1998).  Adopting a low
normalisation for the power spectrum gives a higher value of $\beta=
0.9^{+0.35}_{-0.20}$ (1--$\sigma$). While still consistent 
with most of the estimates above, such large values are only marginally consistent with the measured variance of the IRAS galaxy density field and 
the power spectrum normalization derived by 
Eke, Cole and Frenk (1996). Furthermore, with this normalization the bulk flows are also difficult to reconcile with the observationally favored values of 
Hubble's constant, $h$, and the spectral shape parameter, $\Gamma$. 

A more accurate determination of $\beta$, to 15\% accuracy, is possible by
performing a more detailed comparison between the peculiar velocity field
inferred from the \PZ survey and the measured peculiar velocities at
independent locations. An analysis of this kind is currently in progress.

\section*{Acknowledgements}

EB, LT and CSF thank Michael Strauss and Marc Davis for providing us with
the original version of their reconstruction codes.  The authors wish to
thank Avishai Dekel for providing unpublished bulk velocity data and the
referee, Michael Strauss, for many useful comments and suggestions. This
work was supported by various PPARC grants and by the EC TMR network for
research in ``Galaxy formation and evolution.''  LFAT was supported by a
JNICT 's PRAXIS XXI PhD fellowship. CSF acknowledges a PPARC Senior
Research Fellowship.

\newpage
\section*{}

\begin{table}
\centering

\caption[]{Parameters for the selection function of IRAS Galaxies}
\tabcolsep 2pt
\begin{tabular}{lcccccccc}  \\ \hline
& Sample & $\alpha$ & $\beta$ & $r_s $  & $r_{\star} $ & $l(100)$ &
$\langle n \rangle $ &\\  
\hline
& z-IRAS PSC$z$  & 0.54 & 1.83 & 6.0 & 87.00 & 11.67 & 5.76 10$^-2$ & \\
& R-IRAS PSC$z$  & 0.52 & 1.92 & 6.0 & 90.75 & 11.66 & 4.61 10$^-2$ & \\
& z-IRAS 1.2Jy & 0.49 & 1.80 & 6.0 & 51.27 & 16.21 & 6.41 10$^-2$ & \\
& R-IRAS 1.2Jy & 0.47 & 1.87 & 6.0 & 52.50 & 16.24 & 5.51 10$^-2$ & \\
\hline
\end{tabular}
\label{tab:self1}

\centering
\caption[]{Parameters for the selection function of Abell/ACO clusters} 
\tabcolsep 2pt
\begin{tabular}{cccccccc}  \\ \hline
Sample & $r_{\circ1} $ & $r_{\circ2} $ & $r_{c1} $ &  $r_{c2} $ &
$A_1$ & $A_2$ & $\langle n \rangle $   \\
\hline
Abell/ACO & 31.8 & 44.0 & 180 & 235  & 125 & 289 & 4.61 $\cdot10^{-5}$ \\
\hline
\end{tabular}
\label{tab:self2}

\centering
\caption[]{The cluster/IRAS galaxy relative bias parameter, $b_c$. 
The top and bottom rows give results with and without applying 
the cluster collapsing procedure. 
Column 1: $N_t^{\delta}$, the number of gridpoints used for the 
regression analysis; 
column 2: $N_{d.o.f.}$, the number of independent volumes; 
column 3: $b_c^{\delta}$ from the $\delta$-$\delta$ regression 
and it 1-$\sigma$ error;
column 4: $A_{\delta}$, the zero point offset in the $\delta$-$\delta$ 
regression and its 1-$\sigma$ error;
column 5: $S^{\delta}=\chi^2_{eff.}/N_{dof}$ from the $\delta$-$\delta$ 
regression; 
column 6: $b_c^{\delta}$ from the $v_y$-$v_y$ regression and its 
1-$\sigma$ error;
column 7: $A^{v_y}$, the zero point offset in the $v_y$-$v_y$ regression 
and its 1-$\sigma$ error;
column 8: $S^{v_y}=\chi^2_{eff.}/N_{dof}$ from the $v_y$-$v_y$ regression.}
\tabcolsep 2pt
\begin{tabular}{cccccccc} \\  \hline
$N_t$ & $N_{d.o.f.}$ & $b_c^{\delta}$&
$A^{\delta}$ & $S^{\delta}$ 
& $b_c^{v_y}$&  $A^{v_y}$ & $S^{v_y}$ \\ \hline
6427 & 54 & $ 4.4 \pm 0.6$ & $0.06 \pm 0.07$ & 1.14 & 
$ 4.7 \pm 0.6$ & $-16 \pm 90$ & 1.55  \\ 
6527 & 55 & $ 4.5 \pm 0.6$ & $0.12 \pm 0.07$ & 1.15 &
$ 4.6 \pm 0.6$ & $8.2 \pm 91$ & 1.57  \\ 
\hline
\end{tabular}
\label{tab:pc}
\end{table}

\newpage

\section*{Figure Captions}

\noindent
{\bf Figure 1.} 
Aitoff projection of the galaxy distribution in galactic coordinates, in
the PSC$z$ survey (upper panel) and in a mock catalogue constructed from a
cosmological cold dark matter N--body simulation (lower panel). 
The filled--in regions show
unobserved or obscured regions; the zone-of-avoidance is the
quasi--horizontal strip surrounding the galactic plane.

\noindent
{\bf Figure 2.} Lower Plot: the number of galaxies as a function of 
redshift--distance 
in the PSC$z$ (upper histogram) and 1.2 Jy (lower, shaded histogram)
samples. The curves show the expected counts as a function of distance
estimated from the selection functions. The heavy line at the bottom shows
the predicted distance distribution of Abell/ACO clusters. The labels give
the total number of objects in each sample. Upper plot: the 
ratio between the observed and the expected number of galaxies in 
the \PZ catalogue (thick line) and in the IRAS 1.2 Jy catalogue (thin
line). 

\noindent
{\bf Figure 3.} Mean inter-object separation as a function of radial 
distance (in \hmpc) in the three samples considered in this paper: PSC$z$
galaxies (continuous line), 1.2 Jy galaxies (dashed line) and Abell/ACO
clusters (dot--dashed line).

\noindent
{\bf Figure 4.}  Two dimensional slices through the density and velocity
fields in a mock PSC$z$ catalogue, corresponding to the mock Supergalactic
plane. The hypothetical observer is located in a region analogous to the
Local Group. Data are shown in a sphere of radius 120\hmpc centred on the
observer. Both density and velocity fields have been smoothed using a
Gaussian filter of radius 6 \hmpc. The upper left--hand plot shows the true
fields while the upper plot on the right shows the reconstructed fields
using Method~1. Continuous lines represent isodensity contours with a
spacing of 0.5 in $\delta$. Solid lines encompass overdensities and dashed
lines underdensities. The heavy line indicates the $\delta=0$ contour
level. The amplitude of the velocity vectors is on an arbitrary scale.  In
the lower left--hand plot we show the error map for the density
field. Contours for the absolute value of the discrepancy between
overdensities, $\Delta$, are drawn at steps of $0.25$.  The shaded region
around SGY=0 is the mock zone-of-avoidance. The lower plot on the right
shows the difference between the true and reconstructed velocity fields.

\noindent
{\bf Figure 5.}  Real space density field derived from the PSC$z$ survey.
A slice along the Supergalactic plane is shown.  The field has been
smoothed with a variable Gaussian filter.  The smoothing length is set at a
constant value of 3\hmpc within 30\hmpc and increases linearly with
distance up to a value of 11.25\hmpc at 150 \hmpc, where the most distant
structure are located.  The yellow line shows the $\delta = 0$ contour.

\noindent
{\bf Figure 6.}  Real space density and velocity fields derived from the
PSC$z$ survey. The fields, smoothed with with a 6\hmpc Gaussian are shown
in a slice along the Supergalactic plane.  The most distant structures are
located at 80\hmpc.  The thick continuous line shows the $\delta = 0$
contour. Positive (continuous lines) and negative (dashed lines) contours
are plotted at steps of $\Delta \delta = 0.5$.  The amplitude of the
velocity vectors, obtained for $\beta=1$, is on an arbitrary scale.  This
reconstruction has been performed using Method~1.

\noindent
{\bf Figure 7.}  The same fields as in Fig.~6, but in two slices parallel
to the Supergalactic plane.  The top panels  refer to the slice at
SGZ$=+40$ \hmpc, above the Supergalactic plane, while the ones at the bottom
refer to the slice at SGZ$=-40$ \hmpc, below the galactic plane.  Density
and velocity fields are plotted separately following the same conventions
adopted in Fig.~6.  This reconstruction has been performed using Method~1
for a value of $\beta=1$.

\noindent
{\bf Figure 8.}  The peculiar velocity field reconstructed using Method~1
(top left panel) and Method~2 (top right panel), smoothed with a 6 \hmpc
Gaussian in a slice along the Supergalactic plane. The amplitude of the
velocity vectors is on the same arbitrary scale in the two panels. The
lower panel is a scatter plot of the SGY components of the velocity vectors
illustrated in the upper panels. Only 1000 points, randomly selected from
the ones within 80 \hmpc, have been plotted.  The parameters of the linear
fit (A,B) are indicated in the legend, along with the scatter in the model
velocities ($ \sigma $).

\noindent
{\bf Figure 9.}  The same as Fig.~6 but the density and velocity fields
have been derived from the 1.2Jy survey.  This reconstruction has been
performed using Method~1.

\noindent
{\bf Figure 10.} Map of the difference between the density and velocity
fields obtained from the \PZ and the 1.2 Jy catalogues in a slice along the
Supergalactic plane.  The lower panel shows the difference between the two
overdensity fields, $\Delta=\delta_{PSCz}-\delta_{1.2 Jy.}$: positive
(continuous lines) and negative (dashed lines) isodensity contours are
drawn at steps of $\Delta \Delta=0.25$; a thick line connects the points in
which $\Delta=0$.  The difference between the two velocity fields, both
obtained for $\beta=1$, is displayed in the upper panel.  The amplitude of
the velocity vectors is on an arbitrary scale.

\noindent
{\bf Figure 11.}  Point-by-point comparison of the 1.2Jy and PSC$z$ density
and velocity fields within $80$ \hmpc. Only the values at 1000 randomly
chosen gridpoints are plotted. The upper panel shows a velocity--velocity
scatter plot for the SGY components of the peculiar velocity (assuming
$\beta=1$).  The lower panel shows the $\delta$--$\delta$ comparison.  The
parameters of the linear regression fit are given in both panels.

\noindent
{\bf Figure 12.}  A section, along the Supergalactic plane,
through the real space density and velocity fields derived from the PSC$z$
survey. The fields have been smoothed with a 6\hmpc Gaussian.  This figure
is similar to Fig.~6, except that it displays data in a larger spherical
volume of radius 120\hmpc. The thick continuous line is the $\delta = 0$
contour. Positive (continuous lines) and negative (dashed lines) contours
are plotted at steps of $\Delta \delta = 0.5$.  The amplitude of the
velocity vectors is on an arbitrary scale. This reconstruction has been
performed using Method~1.

\noindent
{\bf Figure 13.}  Four slices, parallel to the Supergalactic plane, through
the same density model illustrated in Fig.~12. The different panels
refer to slices at SGZ=+40, +80, -40 and -80 \hmpc, respectively.
The same conventions as in Fig.~6 are adopted.

\noindent
{\bf Figure 14.} 
PSC$z$  velocity fields corresponding to the density fields of Fig.~13. The
amplitude of the velocity vectors is on an arbitrary scale. 

\noindent
{\bf Figure 15.}  Real space density fields derived from the PSC$z$ survey
(upper panel) and Abell/ACO clusters (lower panel) within a distance of 140
\hmpc.  Both fields have been smoothed with a 20\hmpc Gaussian. Here we
show a slice along the Supergalactic plane.  The $\delta=0$ level is
indicated by the thick line.  Other contours are plotted in steps of
$\Delta \delta = 0.2$ for the PSC$z$ map and $\Delta \delta = 0.88$ for the
clusters map. The dashed lines delineate the zone-of-avoidance in each
sample.

\noindent
{\bf Figure 16.} Cluster (left) and \PZ (right) density fields, as in
Fig.~15, but in four slices along the Supergalactic plane, at
SGZ=+100,+45,-45,-100 \hmpc.  The level of the density contours is also as
in Figure~15.

\noindent
{\bf Figure 17.}  Peculiar velocity fields derived from the PSC$z$ survey
(upper panel) and Abell/ACO clusters (lower panel) corresponding to the
density fields in Fig~15. Both fields have been smoothed with a 20\hmpc
Gaussian.  The amplitude of the velocity vectors is on an arbitrary scale,
but the cluster field has been scaled according to a relative bias
parameter, $b_c=4.4$. The dashed lines delineate the zone-of-avoidance in
each sample.

\noindent
{\bf Figure 18.} Error maps similar to the one displayed in Figure~4, except 
that a different mock \PZ catalogue is used and
a larger Gaussian smoothing of 20 \hmpc is assumed. The density 
contours are drawn at steps of $\Delta \Delta=0.1$ in the upper plots
and of $\Delta \Delta  =0.05$ in the lower left--hand plot.

\noindent
{\bf Figure 19.}  Density and velocity scatterplots for the reconstructions
based on the PSC$z$ survey and a sample of Abell/ACO clusters. For clarity
only $\sim 1000$ out of 1878 gridpoints within a sphere of radius 120\hmpc,
and lying within $|b|\ge 20^{\circ}$, are plotted. The velocity--velocity
comparison in the upper panel refers to the SGY--component.  The lower
panel shows the $\delta$--$\delta$ comparison. The parameters in the legend
refer to the $\chi^2$ fits discussed in the text (see also
Table~\ref{tab:pc}). The errorbars give the mean 1--$\sigma$ errors in the
two model fields.

\noindent
{\bf Figure 20.}  Random and systematic errors in the model bulk flow
derived from the analysis of mock PSC$z$ catalogues. The upper panel shows
the difference between the true cumulative bulk velocity vector and the
vector reconstructed using Method~1.  Different symbols are used for the three
Cartesian components, as indicated in the figure.  The lower panel shows
the ratio between the amplitudes of the reconstructed and true bulk flow
vectors (filled circles) and their SGY Cartesian components (open circles).

\noindent
{\bf Figure 21.}  The amplitude (upper left-hand plot) and the direction
(upper right--hand plot) of the cumulative bulk velocities predicted in
different models. The circles give results from the PSC$z$ survey using
Method~1 (open circles) and Method~2 (filled circles); the filled squares
give results from the 1.2Jy survey using Method~1; the triangles give
results from the Abell/ACO cluster sample. For the models based on IRAS
galaxies, the amplitudes have been normalized to $\beta=1$, while for the
cluster model the velocities have been normalized assuming a relative bias,
$b_c=4.4$. The errorbars give 1--$\sigma$ uncertainties obtained from the
PSC$z$ survey using Method~1. The dot--dashed line shows the PSC$z$
reconstruction using Method~1, statistically corrected to minimise the
volume bias.  The right-hand panel gives the direction of the predicted
cumulative bulk flow at 60 \hmpc in galactic coordinates $(l,b)$. The
asterisk at the centre marks the direction of the CMB dipole and the
different samples are indicated by the same symbols used in the top
left-hand plot. Errorbars are 1--$\sigma$ uncertainties derived from the
analysis of the mock catalogues. The contours are set at constant
misalignment angle from the apex of the CMB dipole, in steps of
$\Delta\theta=10^{\circ}$.  The lower plot shows the misalignment angle
between the various model bulk flows as a function of distance. The
difference angle between models M1 and M2 (continuous line), M1 and 1.2 Jy
(short dashed line) and M1 and Clusters (short dashed line) is displayed.

\noindent
{\bf Figure 22.}  The amplitude (left) and direction (right) of the
cumulative bulk velocity vector. The bottom--left panel refers to the total
velocity and the top--left panel to the SGY--component only. The filled
triangles show the bulk velocity measured from the Mark~III catalogue by
Dekel \etal (1998) and the filled square the bulk velocity measured from
the SFI catalogue by Eldar \etal (1998).  The open triangle shows the bulk
flow inferred from a survey of 44 Type~Ia supernovae by Riess, Press \&
Kirshner (1995). The open square gives the bulk velocity derived by Lauer
\& Postman (1994) from a survey of brightest cluster galaxies.  The
dot--dashed lines bracket the 1--$\sigma$ range of the bulk velocity
predicted from the PSC$z$ gravity field using Method~1, and corrected for
the effects of long-wavelength modes. In the both panels the model
predictions are normalized to $\beta=0.75$.  In the right-hand panel, the
asterisk marks the direction of the CMB dipole in galactic coordinates and
the filled circle the direction of the bulk velocity at 50\hmpc obtained
from the PSC$z$ survey using the corrected Method~1. The other symbols
correspond to those in the left-hand panel, at a distance of 50 \hmpc in
the case of the Mark~III catalogue.

{\bf Figure 23.} Likelihood contours 
of
${\cal L}({\bf v}_M(i),i = 1,4; {\bf v}_{P}(j), j=1,8)$
in the $\beta$, $l_{eq}=1/\Gamma$ plane. 
The minimum (or the minima) is displayed by a small horizontal line.
The left-hand panel assumes $\sigma_8=0.87$ and the right-hand panel
$\sigma_8=0.52$. The vertical lines encompass the range for which
$0.4<h<0.8$, for $\Omega=1$ in the left panel and $\Omega=0.3$ in the right
panel.

{\bf Figure 24.} As Fig.~23 but the contours displayed here 
refer to the conditional likelihood function 
${\cal L}({\bf v}_M(i), i=1,4; {\bf
v}_{P}(j), j=1,8; {\bf v}_d(k), k=1,N; {\bf v}_{LG} | {\bf v}_s)$.

{\bf Figure 25.} Marginal distributions of the likelihood of $\beta$ for
$\sigma_8=0.87$ (left panel) and $\sigma_8=0.52$ (right panel).  Continuous
lines give the result of integrating over all values of $\Gamma$, while
dot dashed lines give the results of limiting the integration to the range
encompassed by the vertical lines in Fig.~24.

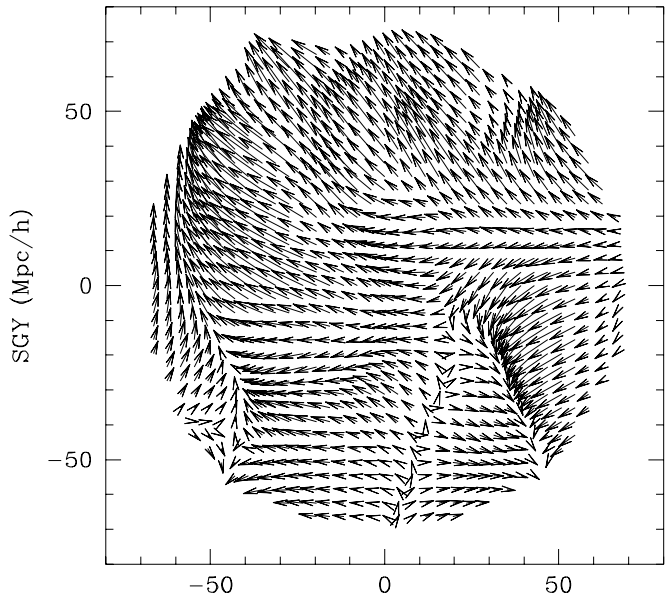
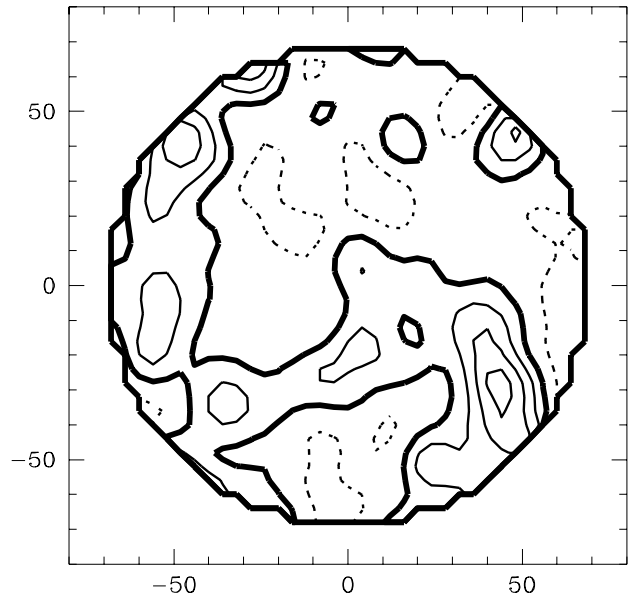
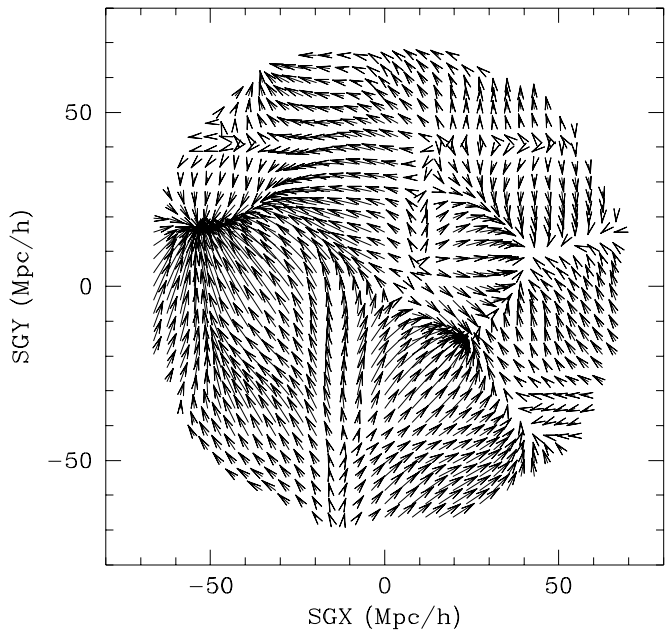
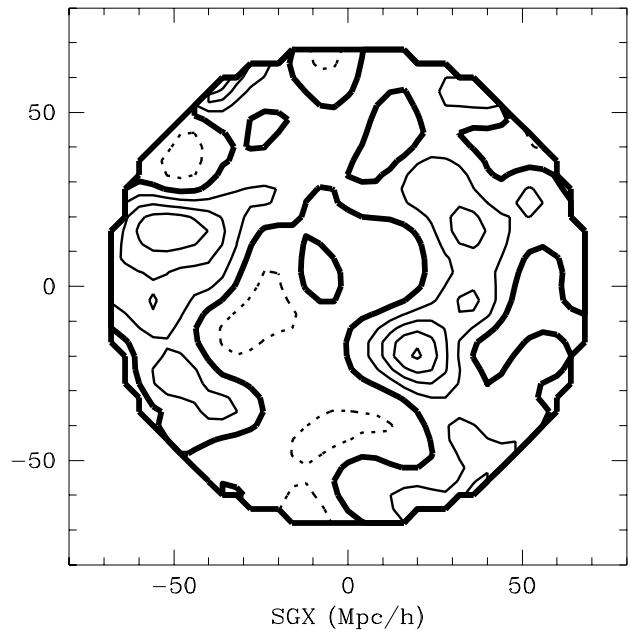
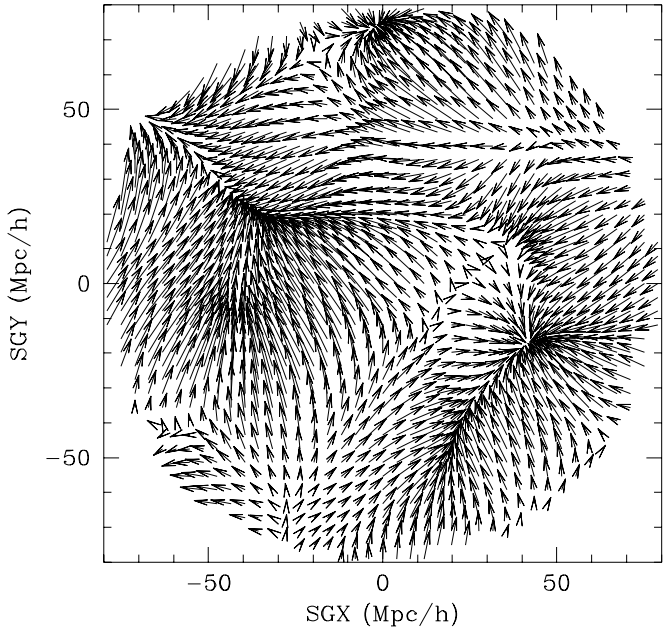
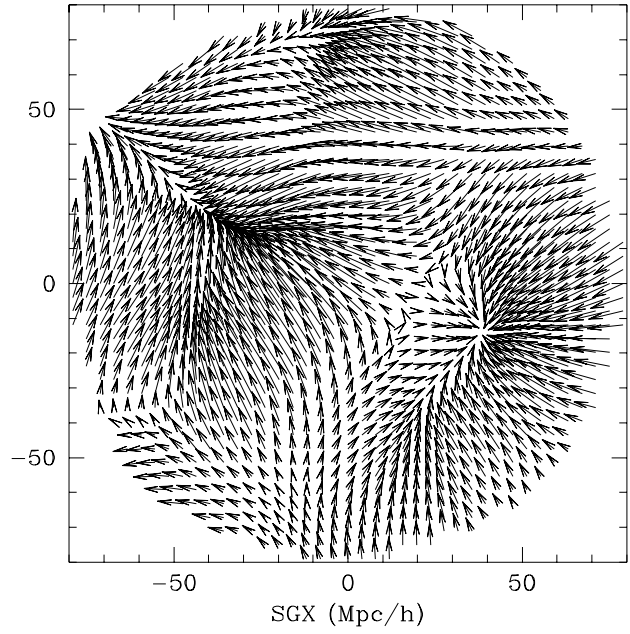
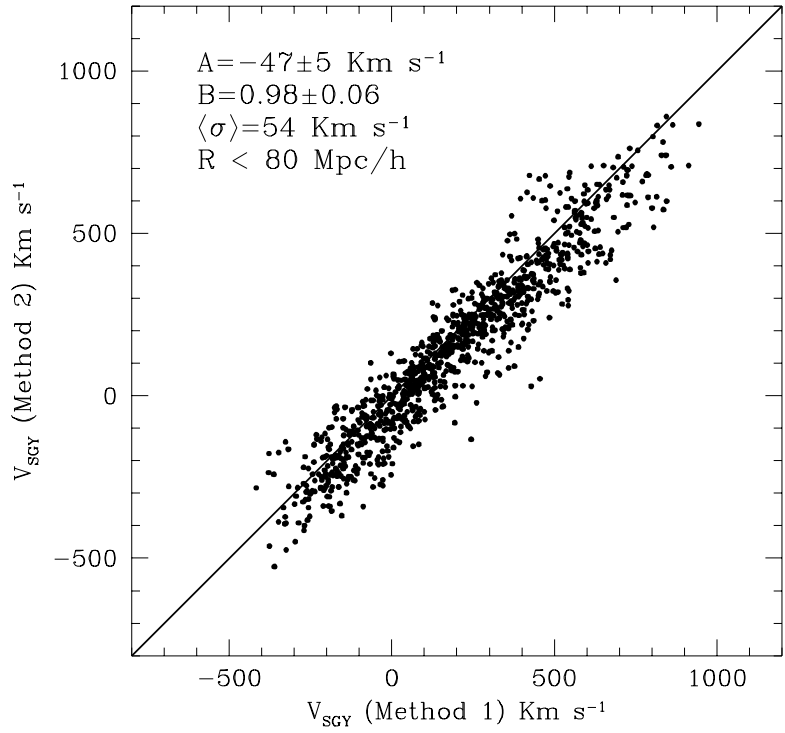
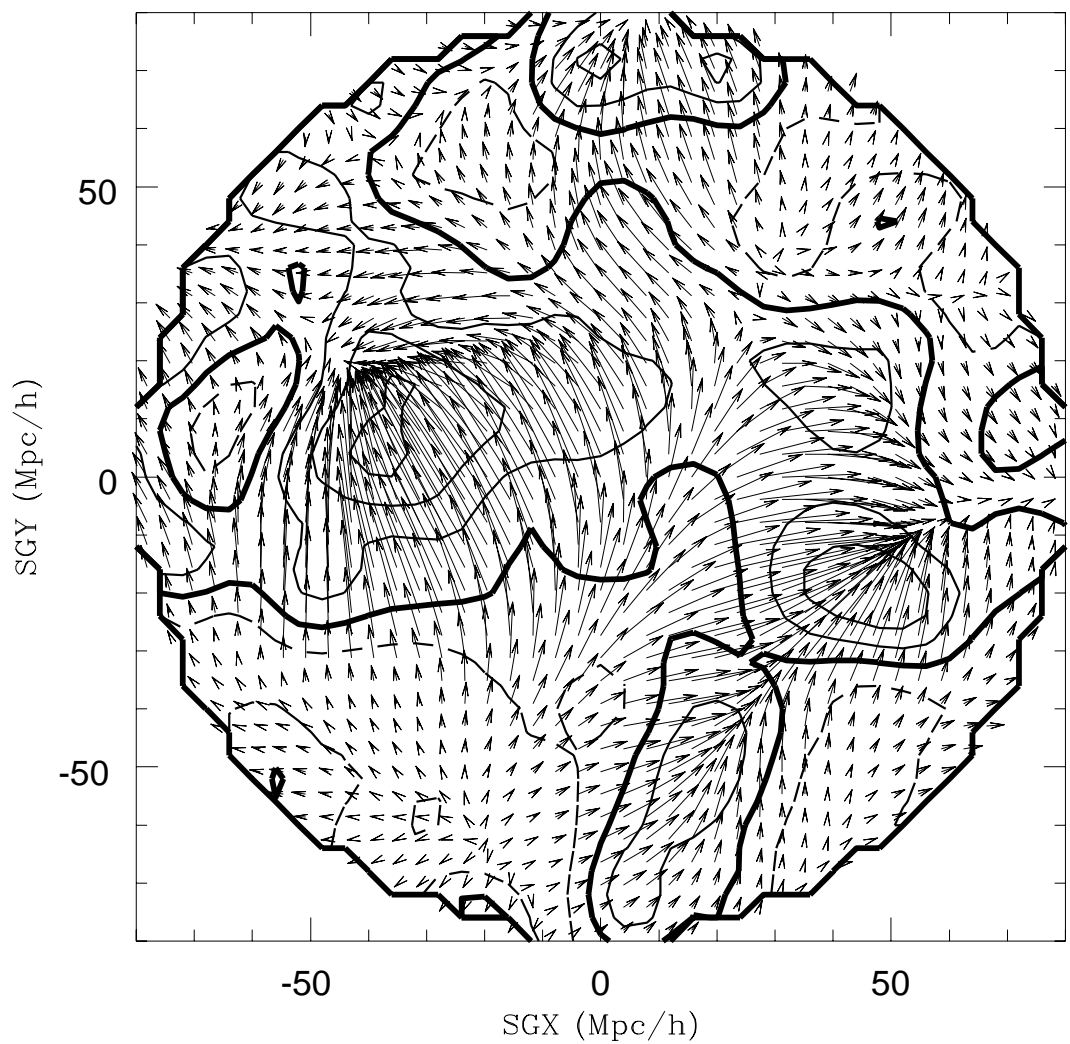
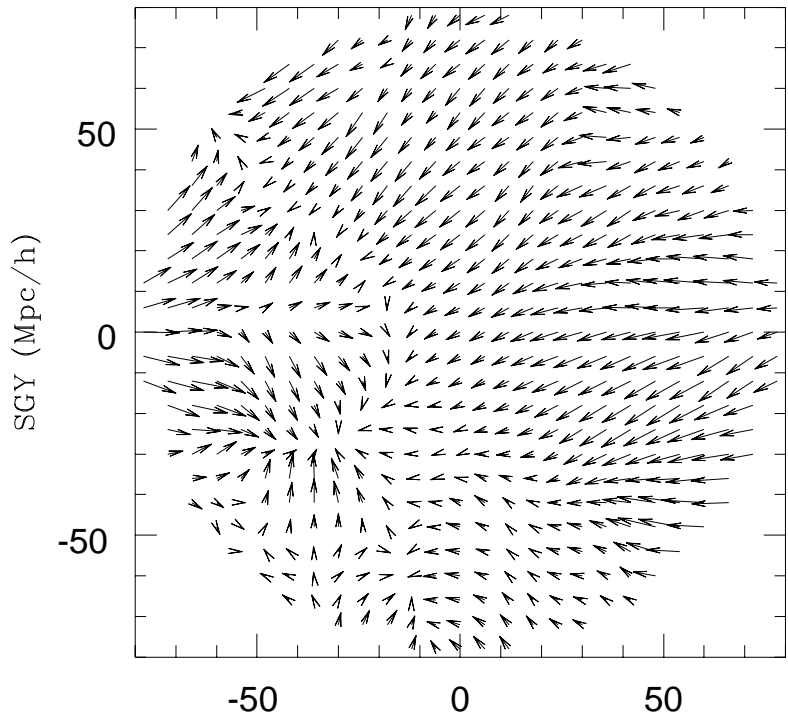
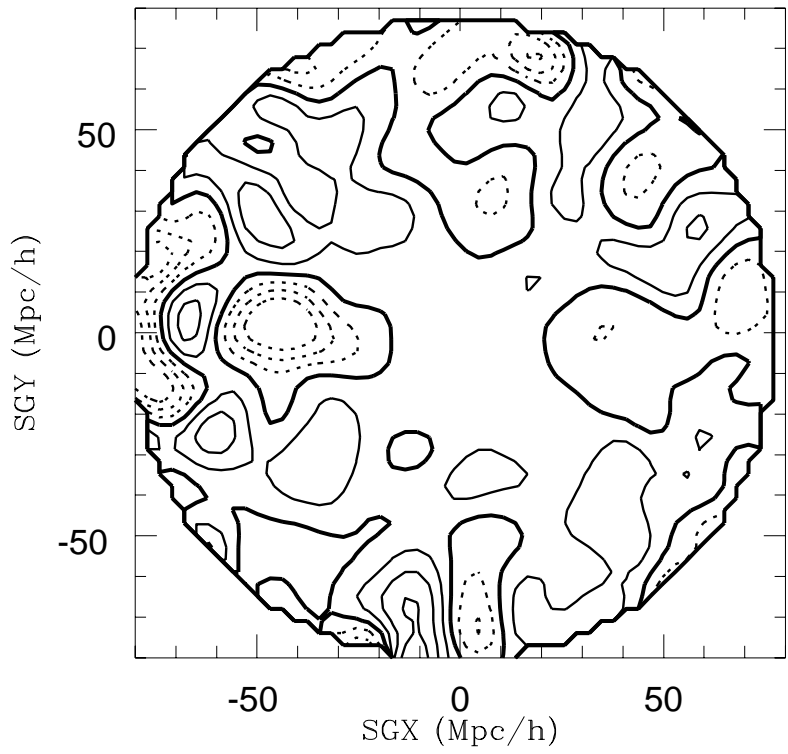
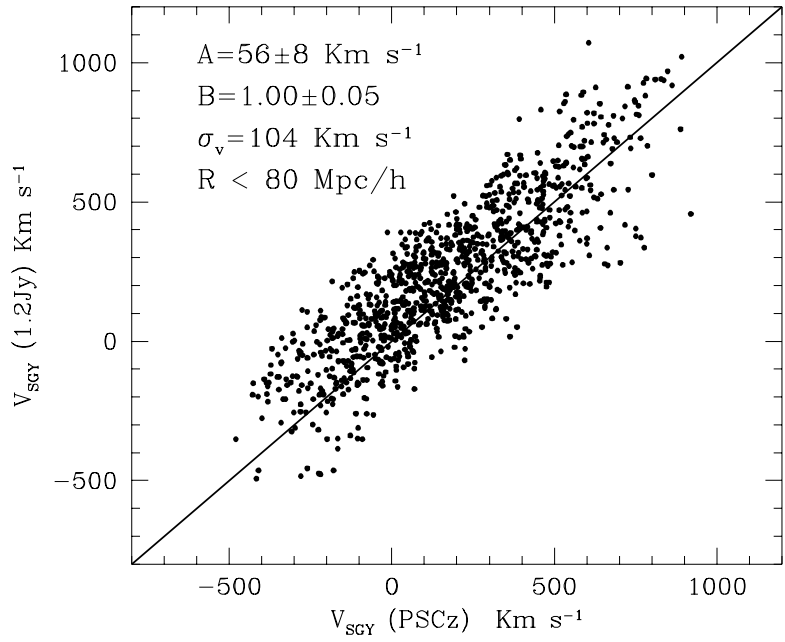
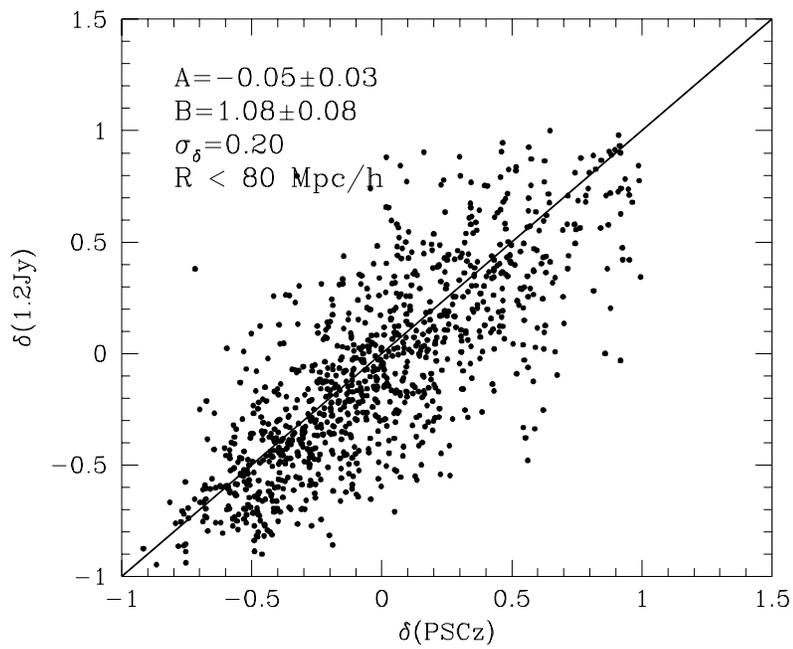
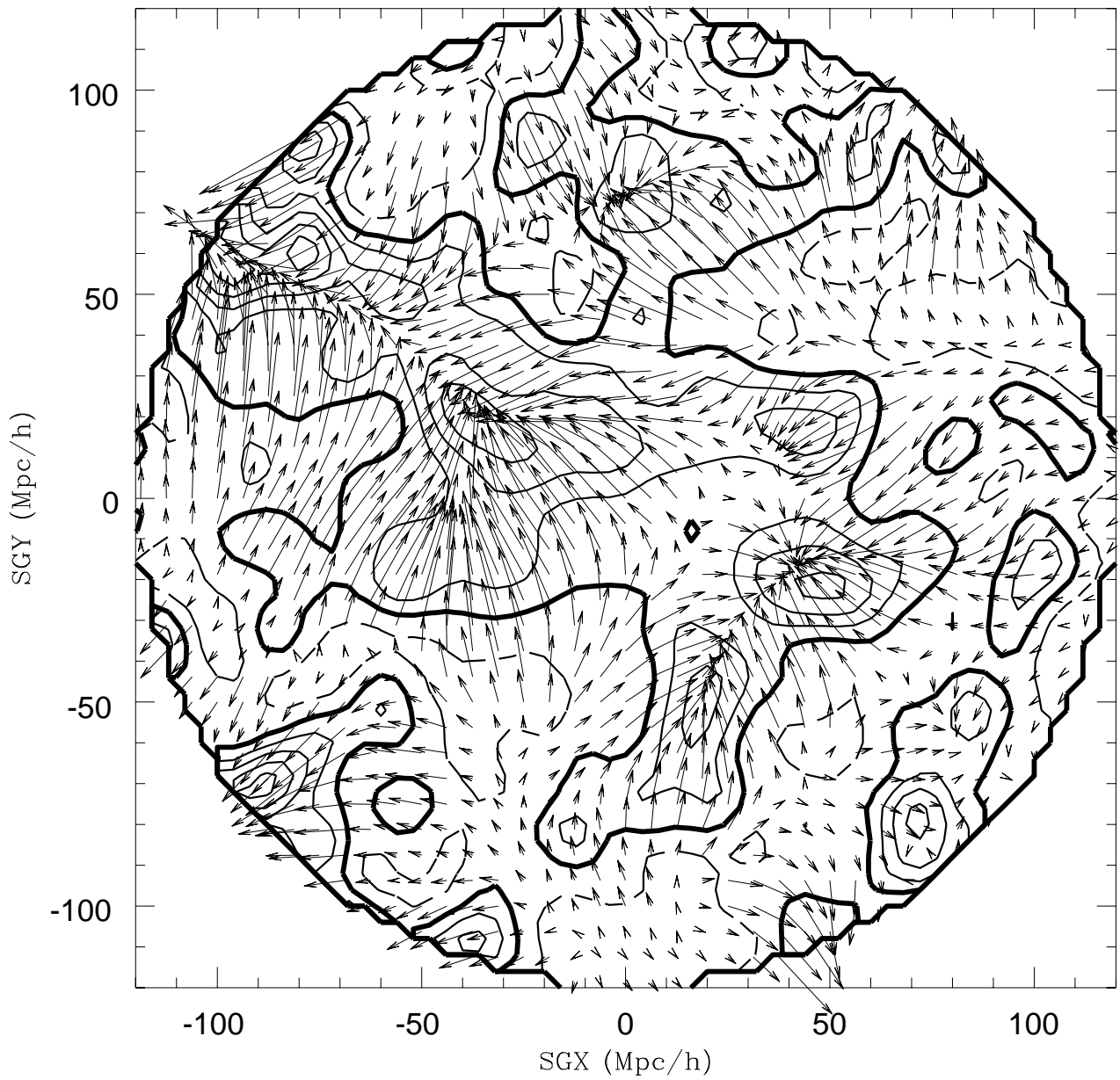
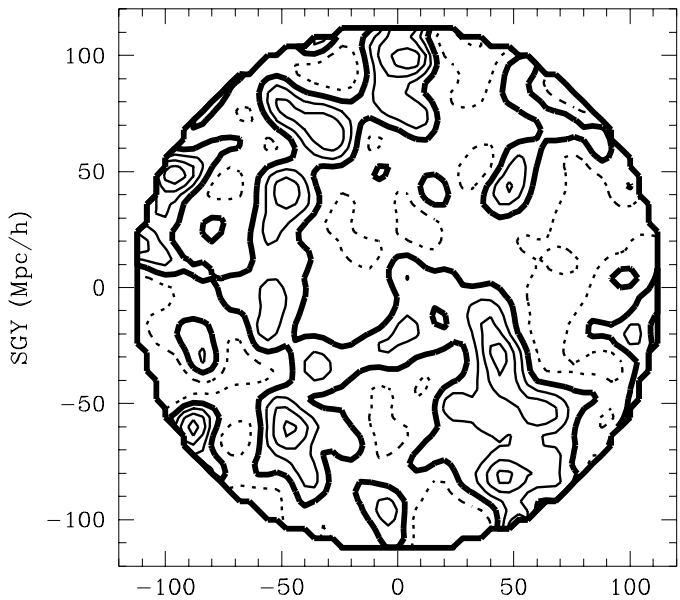
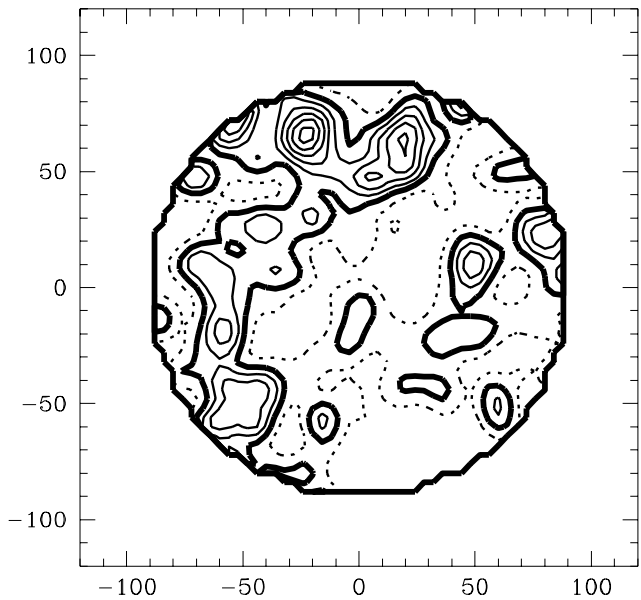
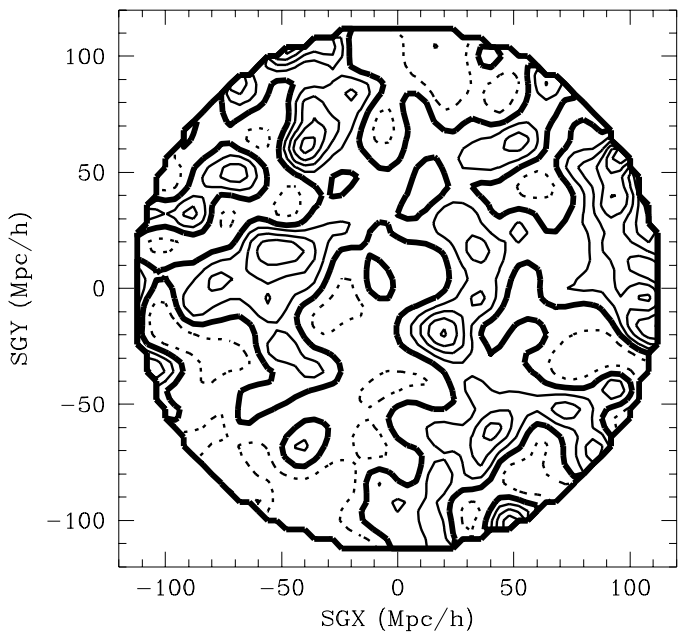
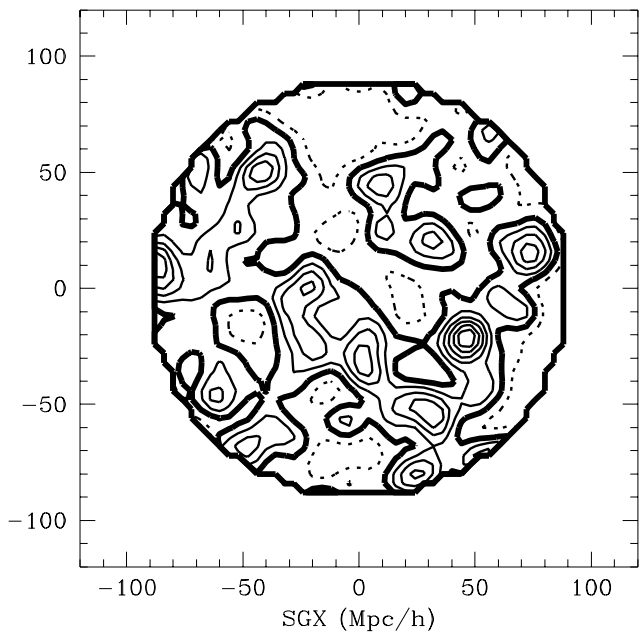
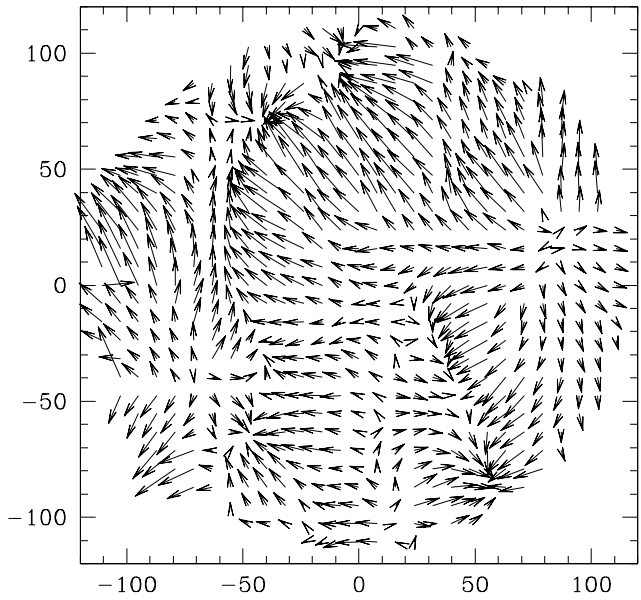
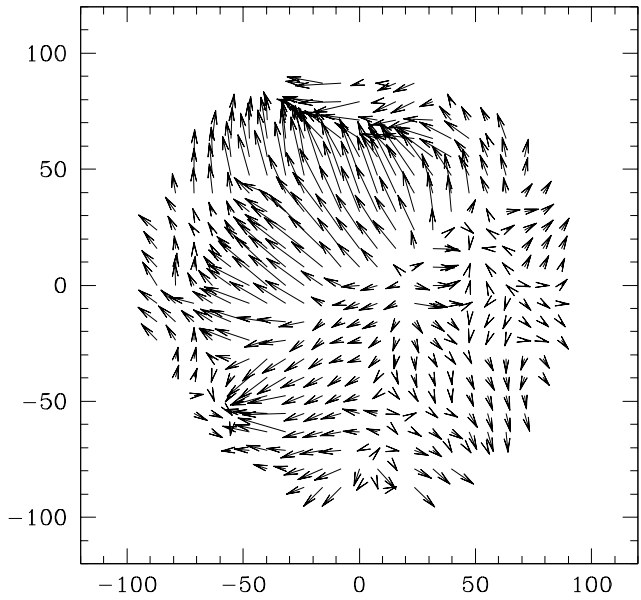
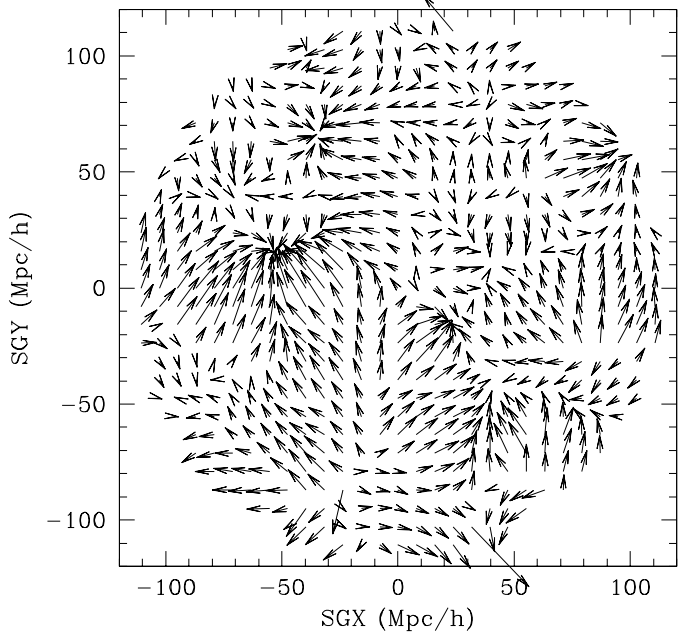
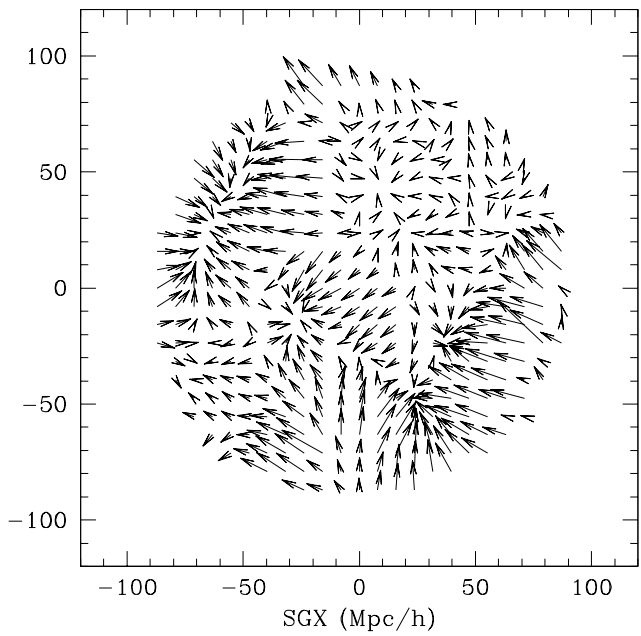
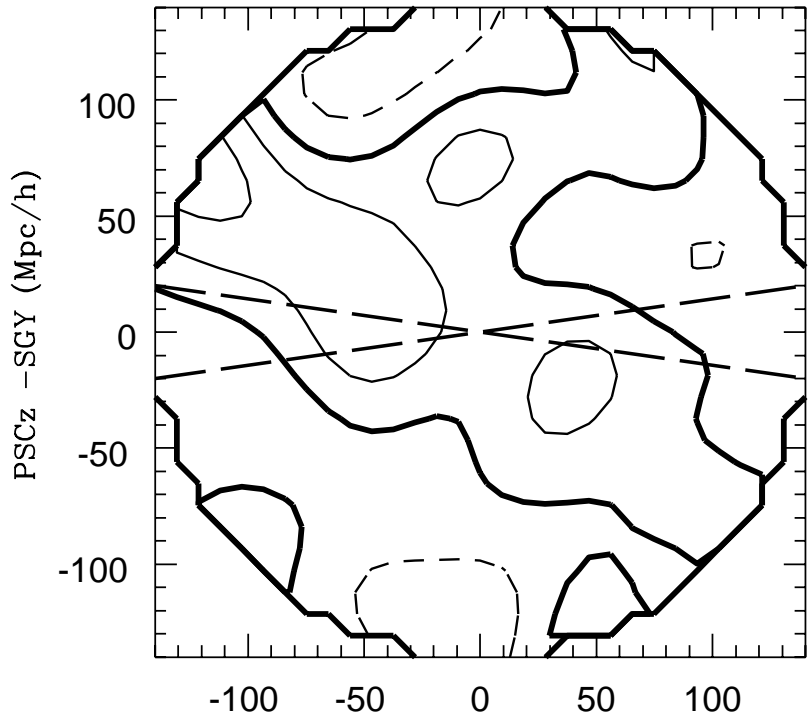
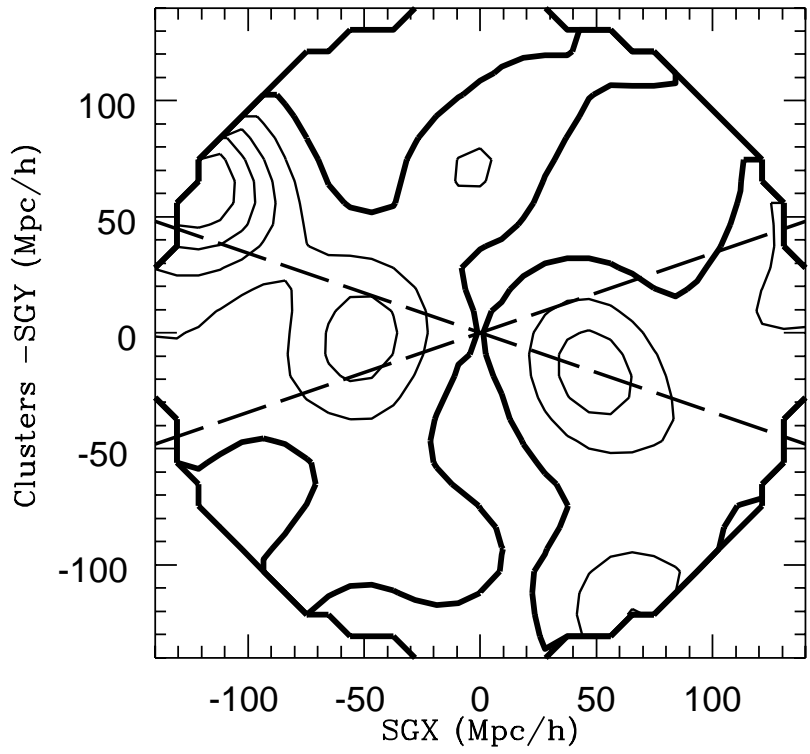
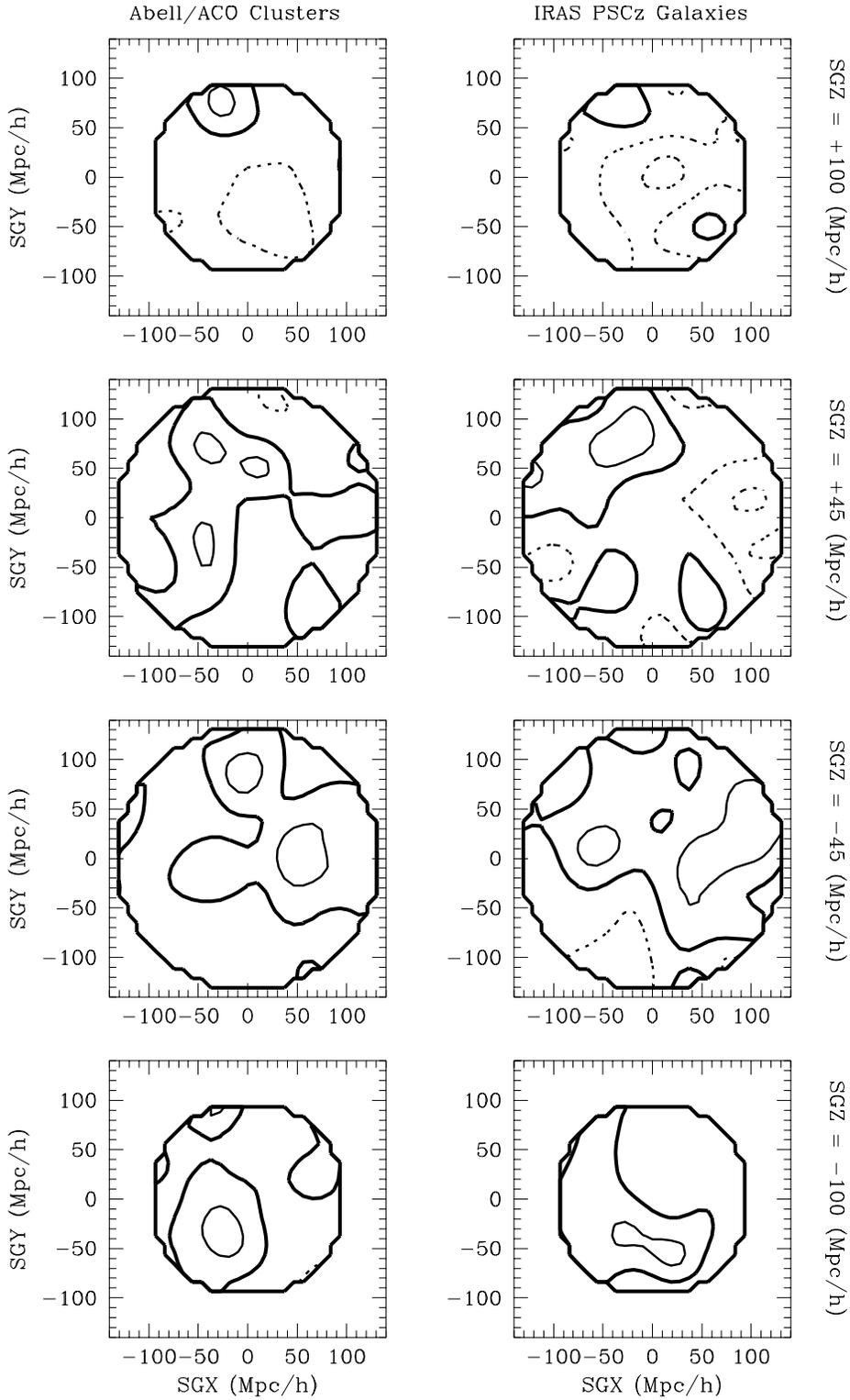
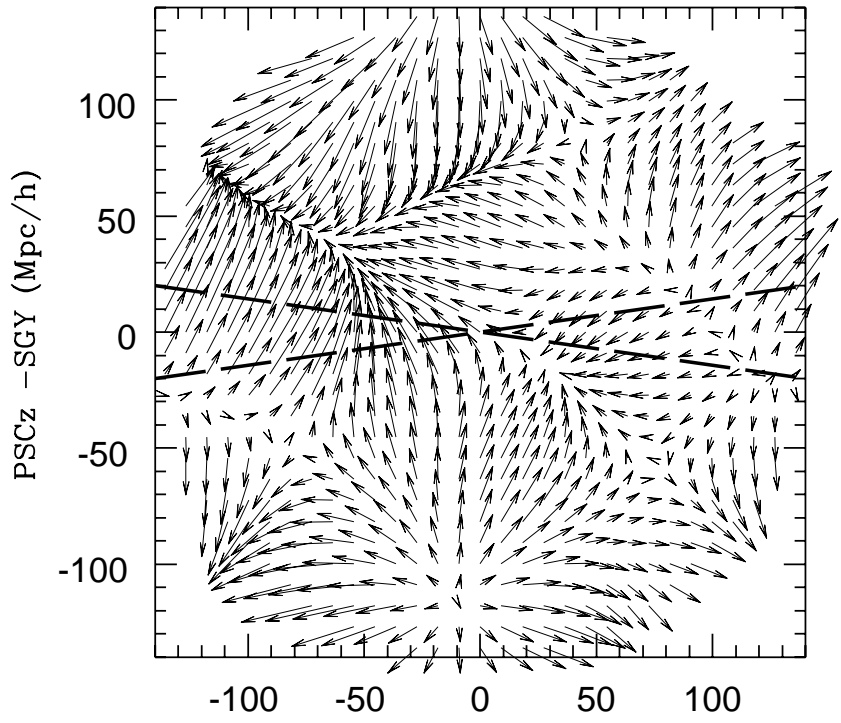
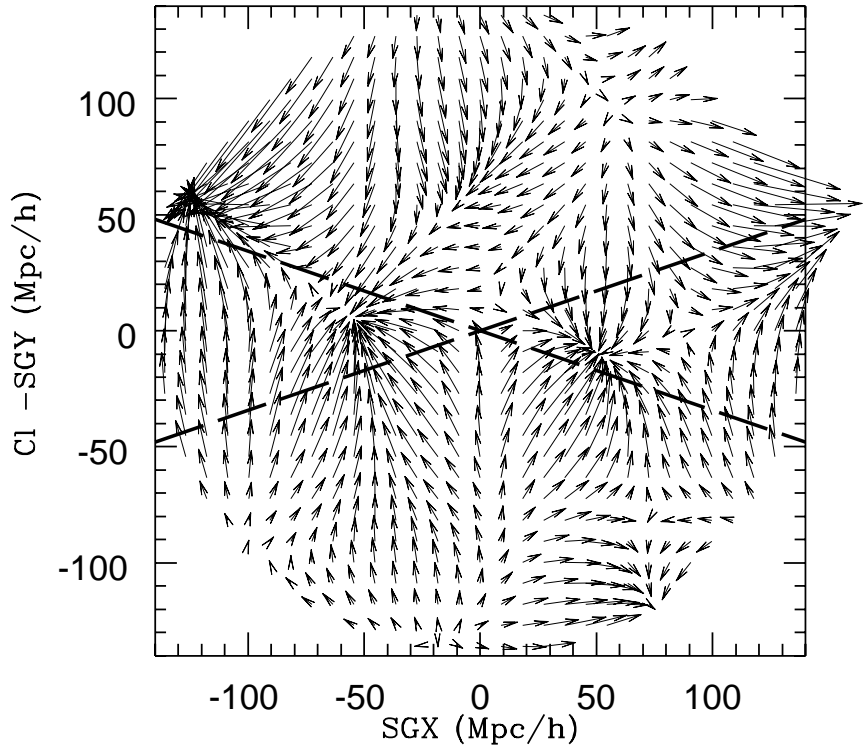
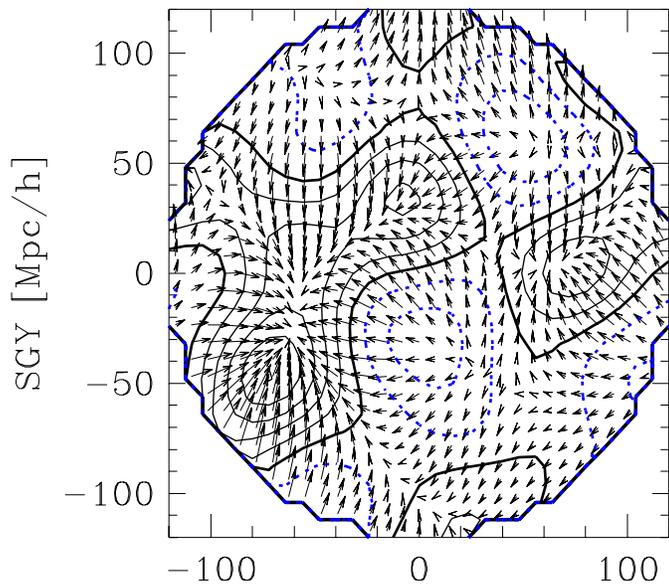
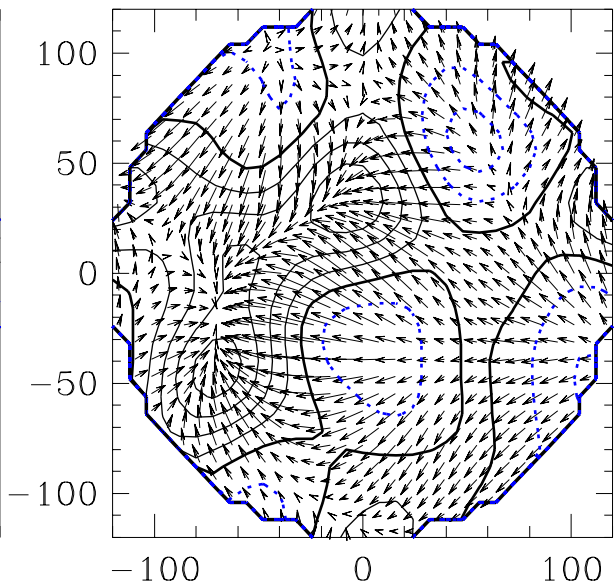
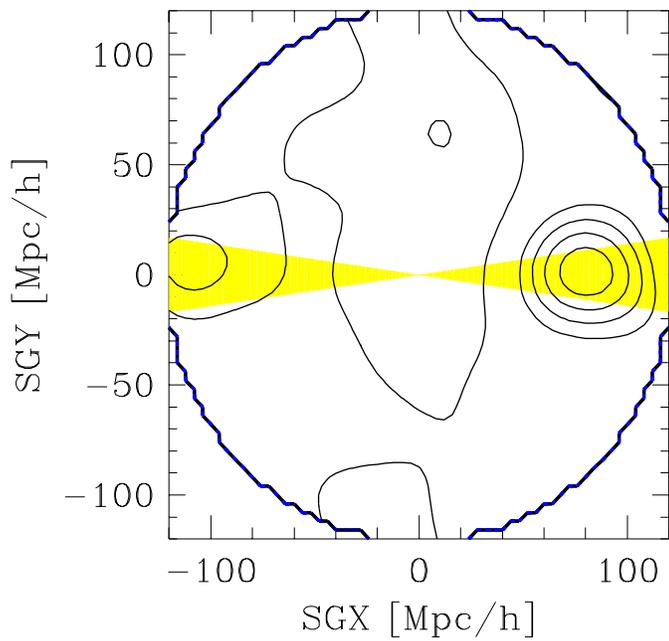
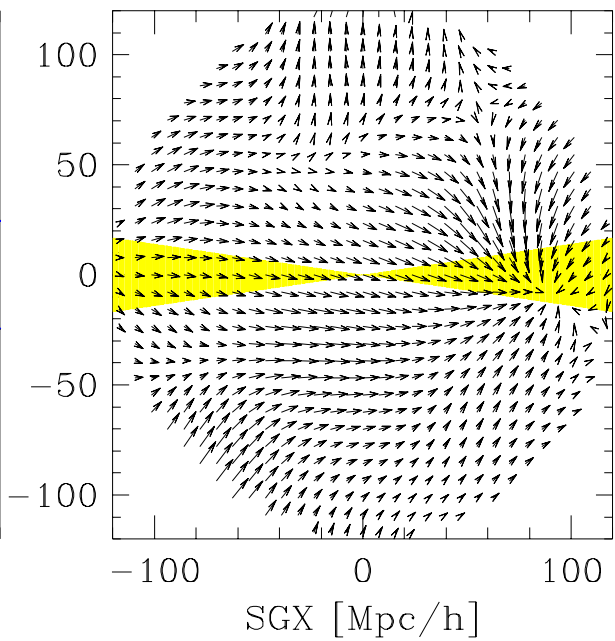
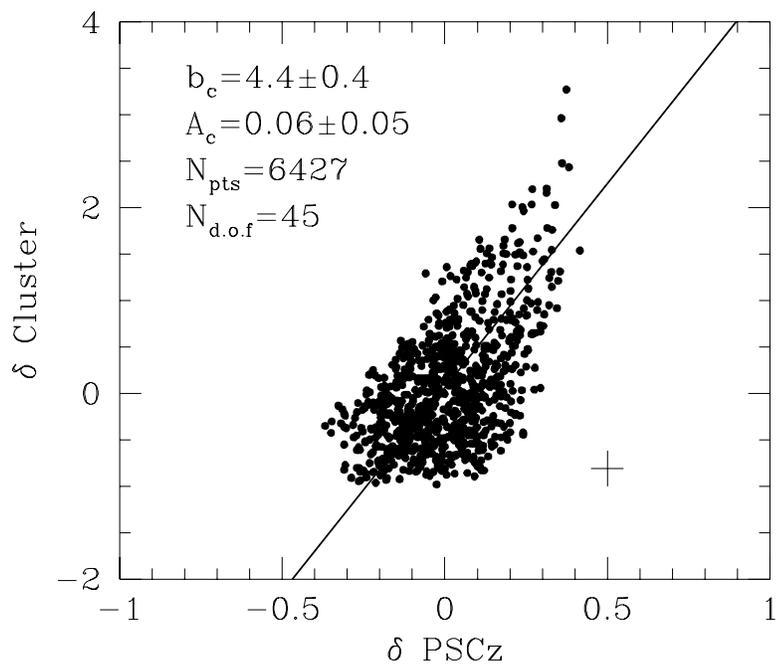
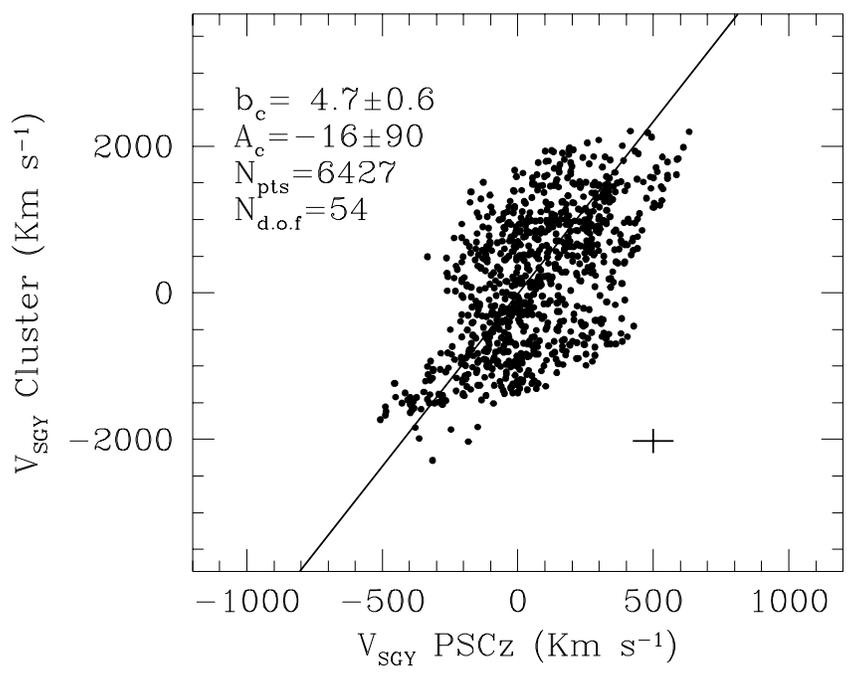
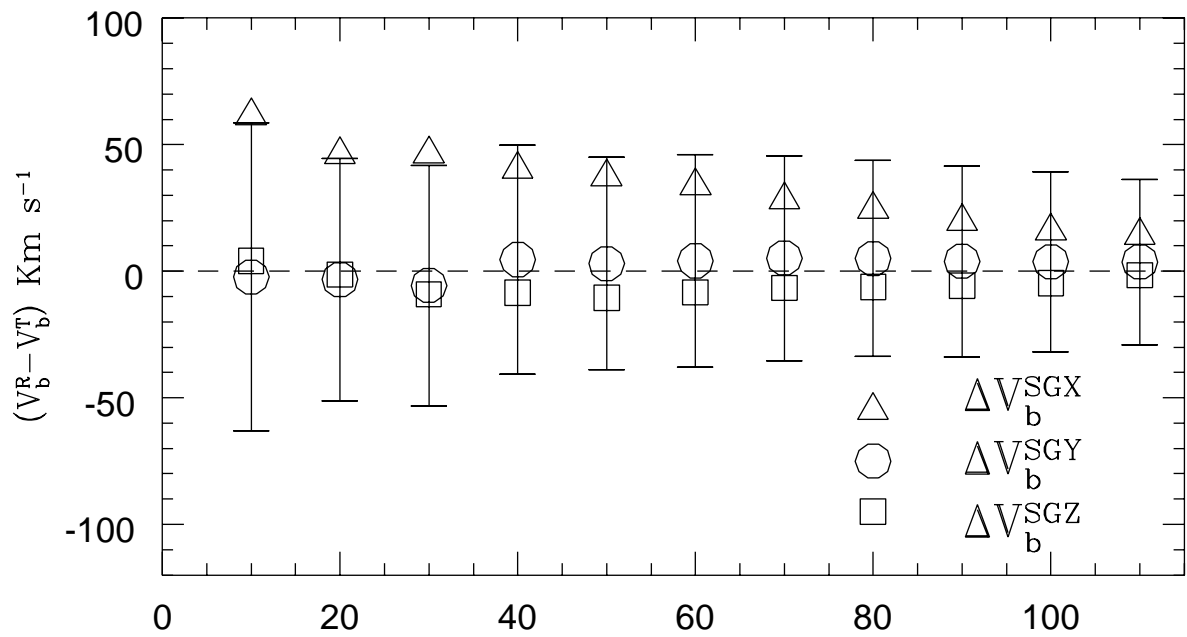
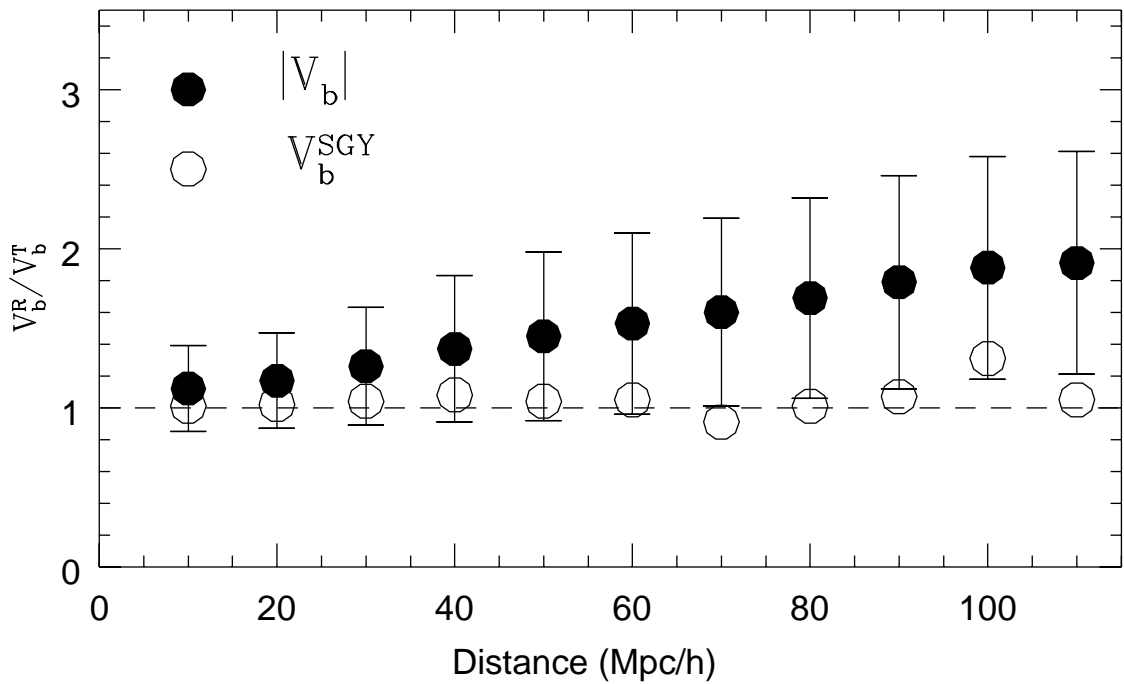
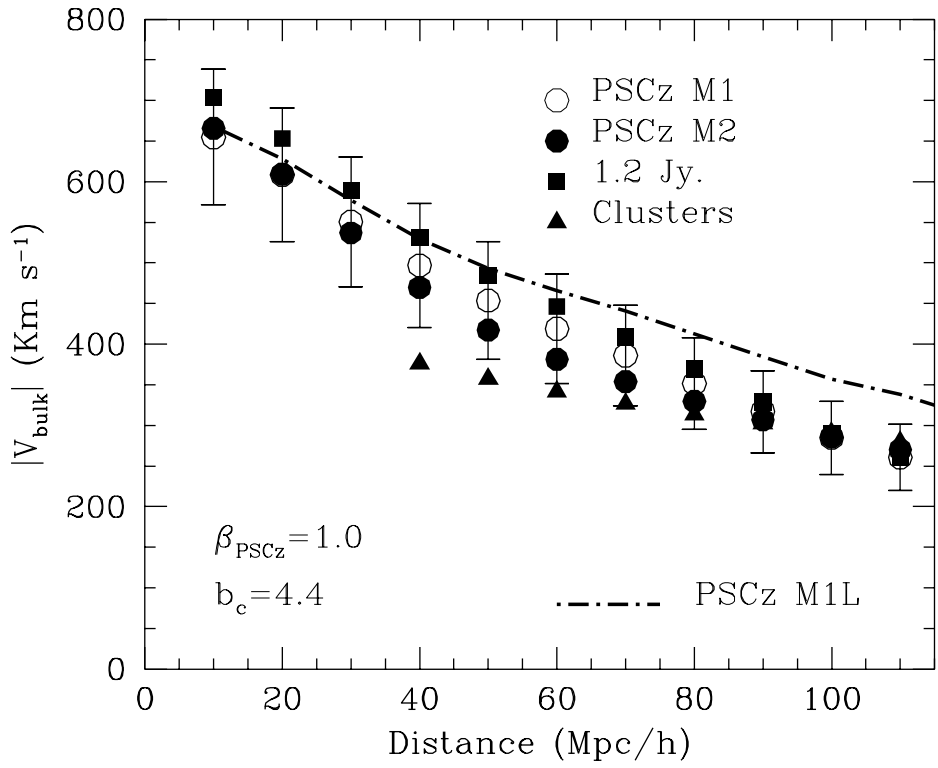
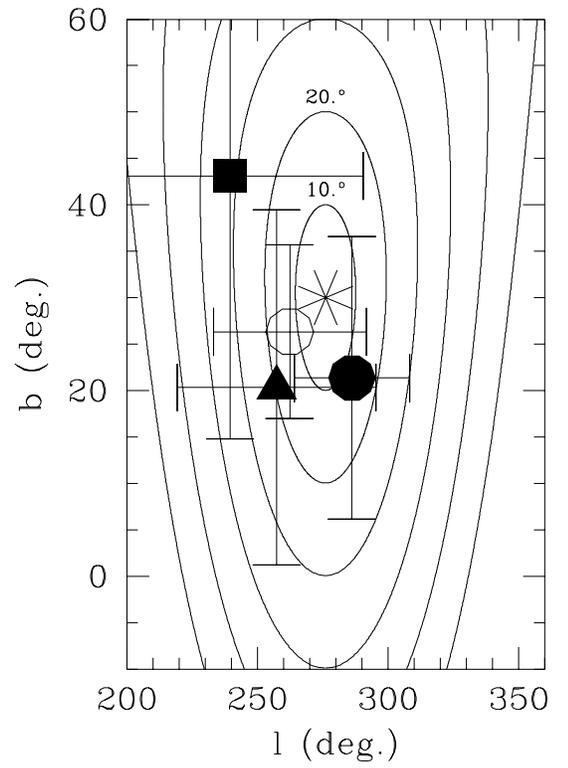
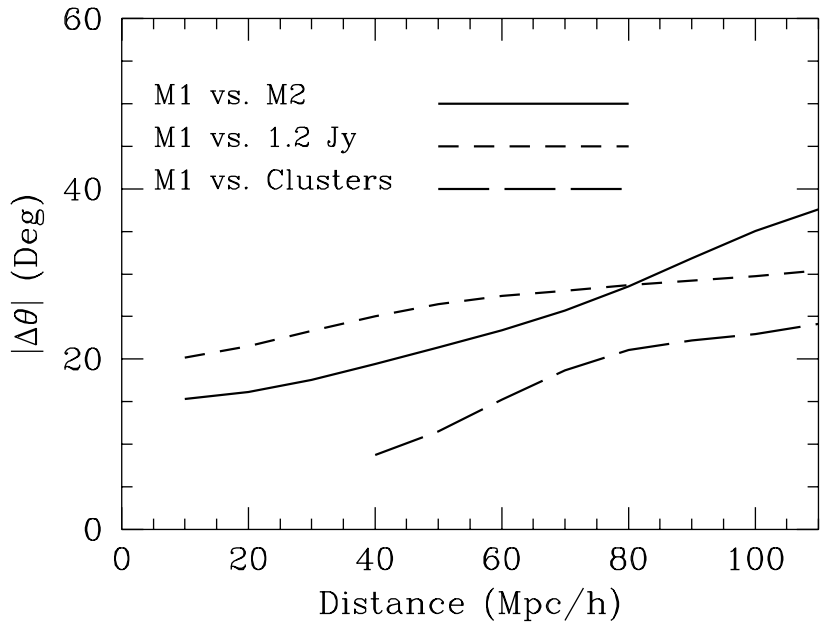
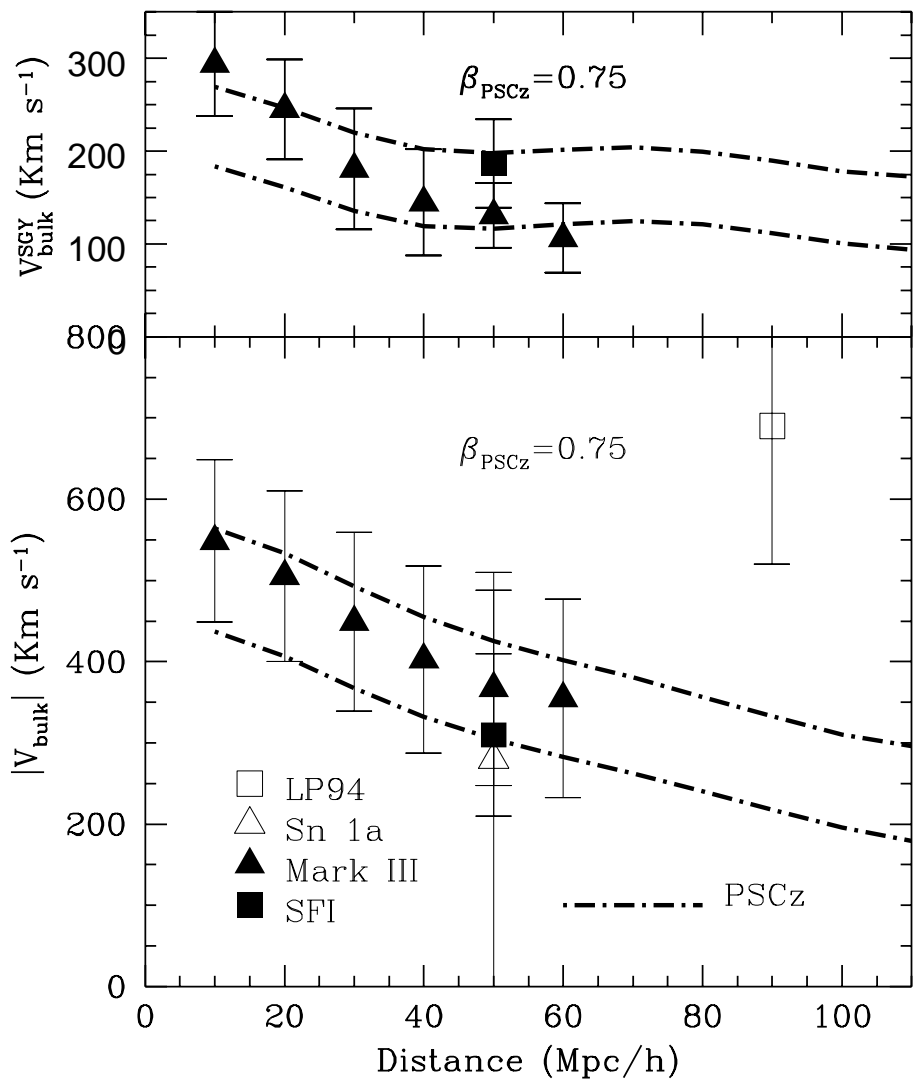
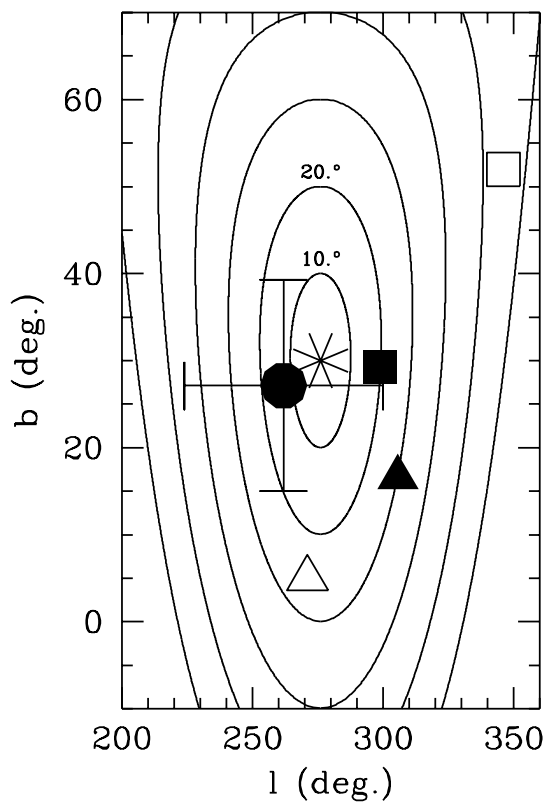
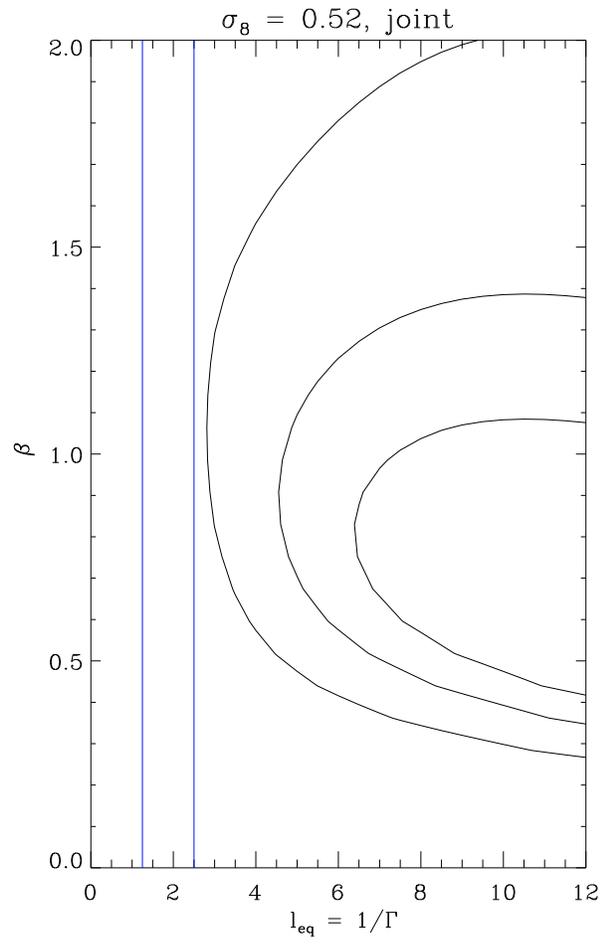
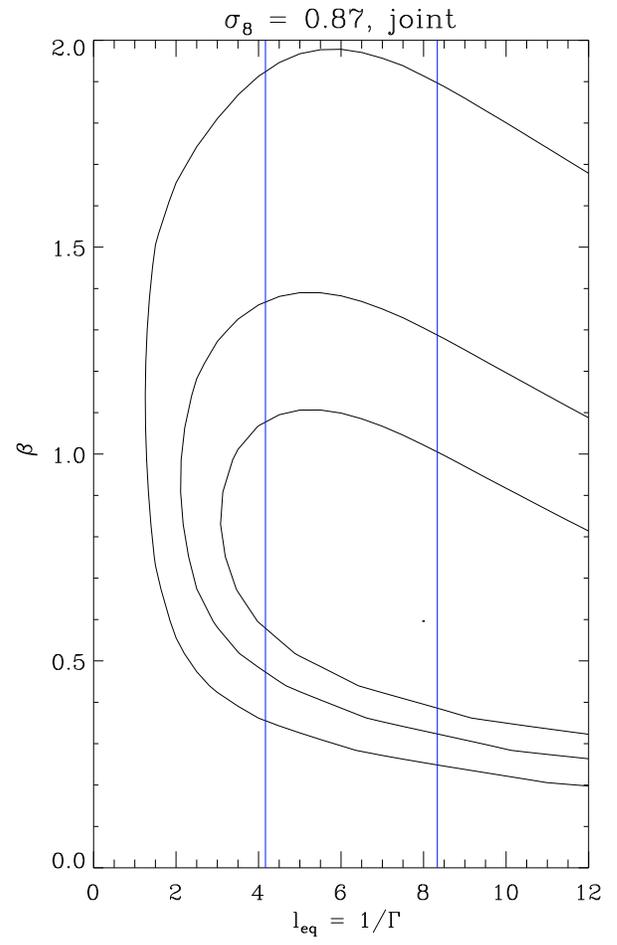
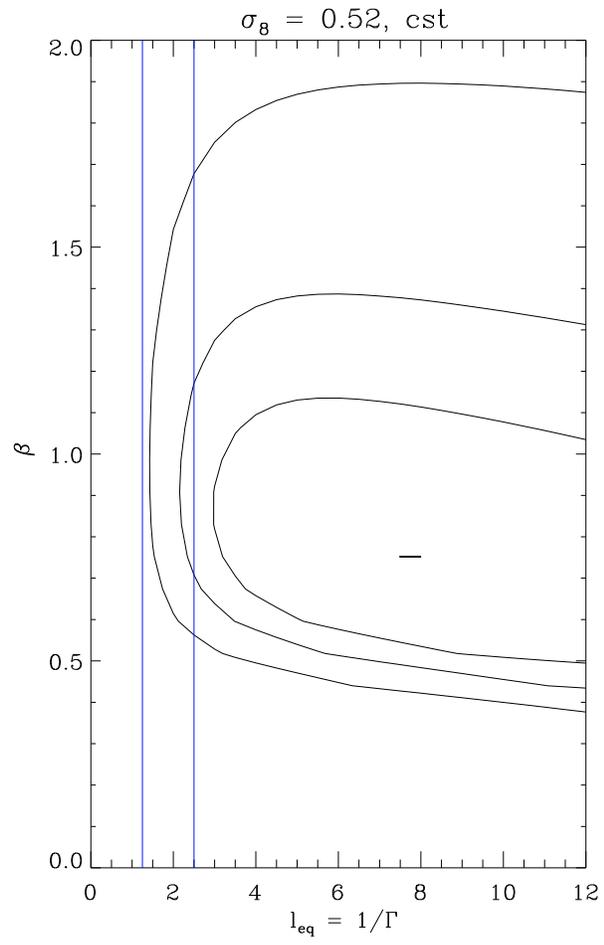
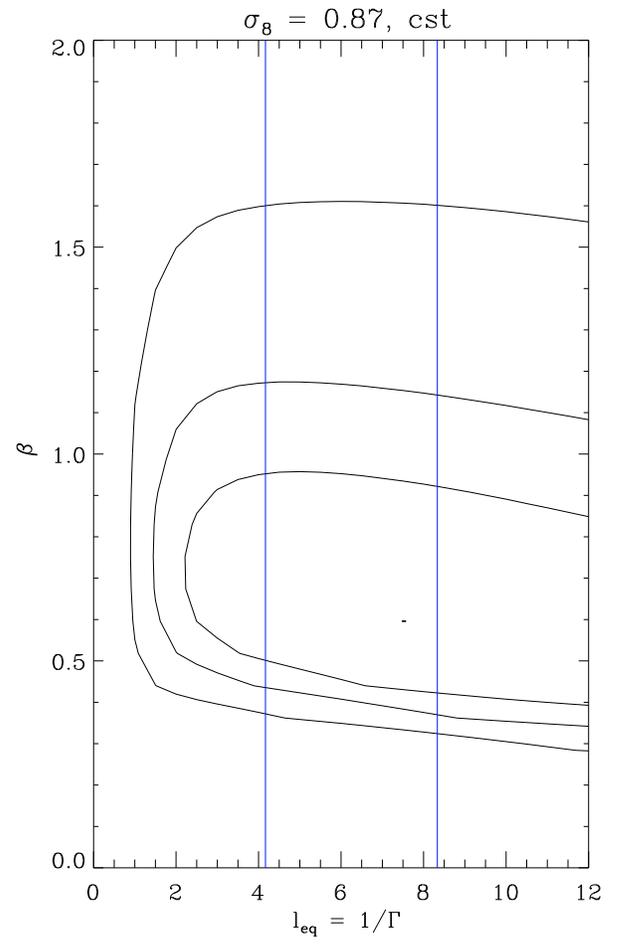
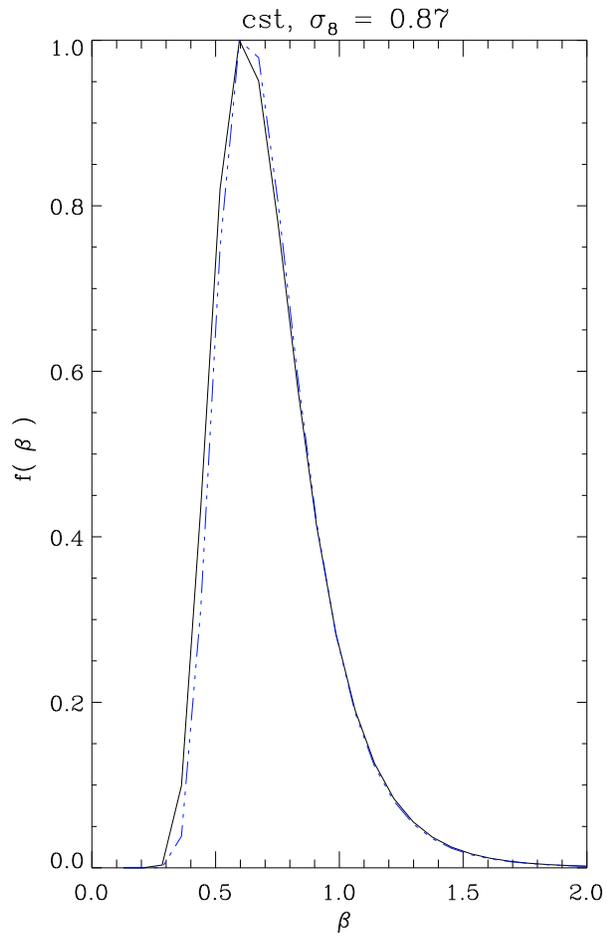
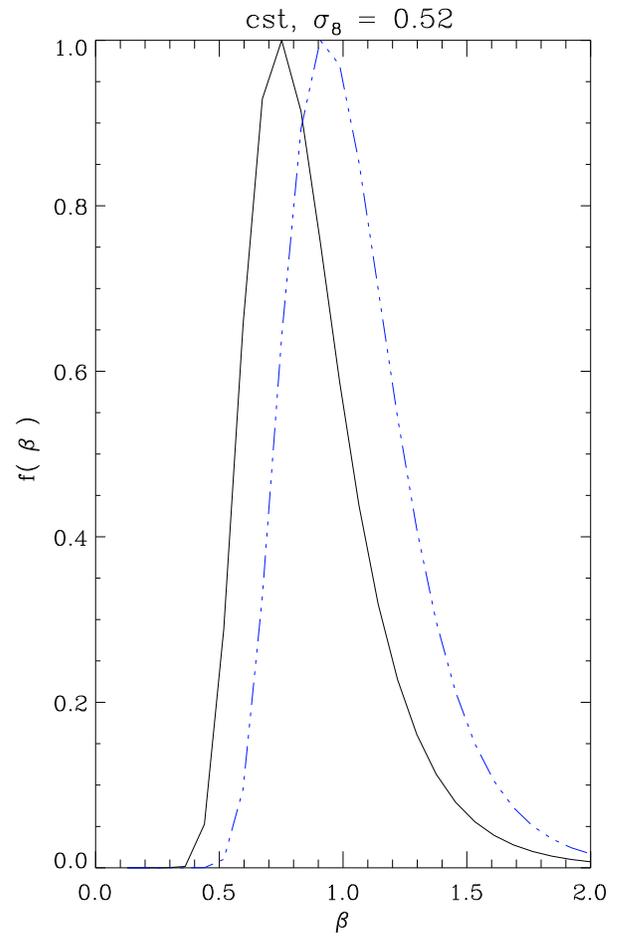
\end{document}